\documentclass[12pt,a4paper]{article}
\usepackage{amssymb}
\input epsf.tex
\usepackage{color}

\def \no {\nonumber}
\def \be {\begin{equation}}
\def \ee {\end{equation}}
\def \bea {\begin{eqnarray}}
\def \eea {\end{eqnarray}}

\def \th {\theta}
\def \p {\partial}

\def\q{q\!\!\!/}

\begin{document}
%\begin{fmffile}{fmfncren}

\baselineskip 0.55 cm
%\begin{flushright}
%SISSA 105/00/EP\\
%hep-th/0011088
%\end{flushright}
\begin{center}

{\Large{\bf Noncommutative QED+QCD and
$\beta$-function for QED}} \vskip .5cm\vskip .5cm

{\Large  M.M. Ettefaghi\footnote{mettefaghi@qom.ac.ir}}
 \vskip .5cm
{\it Department of Physics, Qom University, Qom 371614-6611, Iran}
\vskip .5cm
 {\Large M. Haghighat\footnote{mansour@cc.iut.ac.ir} and
R. Mohammadi\footnote{r.mohammadi@ph.iut.ac.ir}} \vskip .5cm
 {\it Department of Physics, Isfahan
University of Technology, Isfahan, 84156-83111, Iran}

\end{center}

\vskip 2cm
\begin{abstract}
QED based on $\theta$-unexpanded noncomutative space-time in
contrast with the noncommutative QED based on $\theta$-expanded U(1)
gauge theory via the Seiberg-Witten map, is one-loop renormalizable.
Meanwhile it suffers from asymptotic freedom that is not in
agreement with the experiment.  We show that QED part of
$U_\star(3)\times U_\star(1)$ gauge group as an appropriate gauge
group for the noncommutative QED+QCD, is not only one-loop
renormalizable but also has a $\beta$ function that can be positive,
negative and even zero.  In fact the $\beta$ function depends on the
mixing parameter $\delta_{13}$ as a  free parameter and it will be
equal to its counterpart in the ordinary QED for
$\delta_{13}=0.367\pi$.
\end{abstract}
\newpage

\section{Introduction}

The possibility of a noncommutative (NC) space-time coordinates was
arisen  in the string theory \cite{string,NCFT}. Thereafter,  the
theoretical and the phenomenological aspects of the noncommutative
field theory (NCFT) has been extensively studied by many physicists
\cite{NCpapers} .  With respect to the phenomenological point of
view the main question is the value of the NC-parameter that is
about $1 TeV$ in recent works \cite{1tev}. Although these bounds on
the NC-parameter are usually obtained by considering the NCFT at the
tree level, for those works at the higher orders or from the
theoretical point of view the renormalizability of the model is a
crucial question.   The U(n)  gauge field theories in the NC
space-time are one loop renormalizable \cite{bonora} but at the
higher order the  UV/IR mixing causes such theories to be
nonrenormalizable \cite{nonre, hayakawa}.  Meanwhile to avoid the
UV/IR mixing, one can construct a $\theta$-expanded NC-quantum field
theory  via the Seiberg-Witten map \cite{ex-nc}.  Although the pure
gauge sector of such a theory is renormalizable \cite{renorm-exnc},
 it is shown that in general the noncommutative QED cannot be
renormalized by means of the Seiberg-Witten expansion even at the
one loop \cite{nonre-exnc}. Therefore it seems that
$\theta$-unexpanded NCQED is more safe for the phenomenological
purposes.  Unfortunately in this case, even in the vanishing limit
of the parameter of noncommutativity, one encounters an asymptotic
freedom for QED   \cite{hayakawa, sadooghi} where the increasing
coupling  with  increasing energy is well tested experimentally.
Furthermore, for the $\theta$-unexpanded gauge groups in the NC
space-time one meets \cite{no-go}:
\begin{itemize}
  \item Only the U(n) gauge groups in the NC space-time ( $U_\star(n)$ ) are allowed.
  \item The fundamental and antifundamental representation of $U_\star(n)$ can only be
  occurred. For example,
  in the  $U_\star(1)$ case, for
arbitrary fixed charge q only the matter fields with charges $\pm q$
and zero are permissible \cite{hayakawa,no-go}.
  \item When we have a gauge group constructed by the direct product
  of several simple gauge groups, a matter field can be charged only under
  two of them.
\end{itemize}
Therefore, for instance,  to construct the standard model of the particle physics (SM) with $SU(3)\times SU(2)\times
U(1)$  as the gauge group containing the quarks, one needs additional consideration.  It is shown in reference \cite{NCSM}
how one can consistently reduce $U_\star(3)\times U_\star(2)\times U_\star(1)$ to the usual SM gauge group through a two-step
appropriate Higgs mechanism.  The negative $\beta$-function for NCQED was obtained for matter fields with $\pm1$ and zero
charges besides pure QED interaction.  But it is well known that in the NCFT  different gauge fields not only have self
interactions ( three and four photon vertices in NCQED) but also they can have interactions among each other (for instance
gluon-photon interaction).   Therefore one can ask if the  $\beta$-function for NCQED after the full consideration is really negative?
In fact to answer to this question we need to consider all particles (leptons as well as quarks) and all gauge bosons.
For this purpose to obtain the $\beta$-function for  QED we explore the quantization of  QED+QCD in the NC space-time and
ignore the weak interaction for simplicity.

This paper is organized as follows: In Sect. 2 we give a brief review on
the $U_\star(3)\times U_\star(1)$ model.  In Sect. 3 we explore the
quantization of the model and then we calculate the QED
$\beta$-function in Sect. 4.  We summarize our results in Sect. 5.
%=========================================================================

\section{ $U_\star(3)\times U_\star(1)$ model}

In this section we review briefly the $U_\star(3)\times U_\star(1)$
model which is introduced in \cite{NCSM}. The $U_{\star}(3)\times
U_{\star}(1)$ theory is described by one gauge field
valued in the $U_{\star}(1)$ algebra, $B_{\mu}$, and  $U_{\star}(3)$-valued
gauge fields as
                \be G_{\mu}(x)=\sum G^{A}_{\mu}T^{A},\label{1}\ee
in which the generators $T^{a} ,a=1,\dots ,8$, are the Gell-Mann
matrices and  $T^{\circ}=(1/6)^{1/2}\times I_{3\times3}$. The finite
local transformations of the gauge fields are :
    \bea B_{\mu}=V\star B_{\mu}\star V^{-1} +
    i/g_{1}V\star \partial_{\mu}V^{-1},\nonumber\\
     G_{\mu}=U\star G_{\mu}\star U^{-1} +
     i/g_{3}U\star \partial_{\mu}U^{-1}\label{2},\eea
where $V\in U_\star(1)$ and $U\in U_\star(3)$ and the  $\star$-product  is defined as:
\be f\star
g(x,\theta)=f(x,\theta)\exp(\frac{i}{2}\overleftarrow{\partial}_\mu
\theta^{\mu\nu}\overrightarrow {\partial}_\nu)g(x,\theta),
\label{a3}\ee where $\th^{\mu\nu}$ is an antisymmetry tensor that
denotes the noncommutativity of space time through \be
[x^\mu,x^\nu]=i\th^{\mu\nu}.\ee  Then the invariant action of the $U_{\star}(3)\times U_{\star}(1)$
Yang-Mills theory is
    \be S_{NCYM}=-1/4\int d^{4}x (B_{\mu\nu} B^{\mu\nu}+ Tr(G_{\mu\nu}
    G^{\mu\nu})),\label{3}\ee
   in which
     \bea B_{\mu\nu}=\partial_{[\mu} B_{\nu]} +
    ig_{1}[B_{\mu},B_{\nu}]_{\star},\nonumber\\
   G_{\mu\nu}=\partial_{[\mu} G_{\nu]} +
    ig_{3}[G_{\mu},G_{\nu}]_{\star}\label{4}.\eea
The independent matter fields that can be accommodated in this model
are electron in the anti-fundamental representation of
$U_{\star}(1)$,  up quark in the fundamental representation of
$U_{\star}(1)$ and in the anti-fundamental representation of
$U_{\star}(3)$, and down quark in the anti-fundamental
representation of $U_{\star}(3)$. Meanwhile the electron neutrino can be
accommodated in the adjoint representation of $U_{\star}(1)$. Thus
the gauge transformation properties of the fermions are
\bea\Psi_{e}(x)&\rightarrow& \Psi_{e}(x)\star V^{-1}(x),\nonumber \\
\Psi_{\nu_e}(x)&\rightarrow& V(x)\star\Psi_{\nu_e}(x)\star V^{-1}(x),\nonumber \\
\Psi_{u}(x)&\rightarrow&  V(x)\star \Psi_{u}(x)\star U^{-1}(x),\nonumber \\
\Psi_{d}(x)&\rightarrow& \Psi_{d}(x)\star U^{-1}(x).\label{5}\eea The
gauge invariant action of this model is
 \bea S_{NCYM}=\int d^{4}x[\hspace{-.6cm}&&\bar{\psi_{e}}\star\gamma^{\mu}D_{\mu}^{1}\psi_{e}
 +\bar{\psi_{\nu_e}}\star\gamma^{\mu}D_{\mu}^{1}\psi_{\nu_e}+\bar{\psi_{u}}\star\gamma^{\mu}D_{\mu}^{1+3}\psi_{u}
 + \bar{\psi_{d}}\star\gamma^{\mu}D_{\mu}^{3}\psi_{d}\no\\ \hspace{-.6cm}
 && - 1/4(B_{\mu\nu} B^{\mu\nu}+ 2Tr(G_{\mu\nu}
    G^{\mu\nu}))],\label{6}\eea
in which \bea D_{\mu}^{1}\psi_e&=&\partial_{\mu}\psi_e-\frac i 2 g_{1}\psi_e\star B_{\mu},\nonumber \\
D_{\mu}^{1}\psi_{\nu_e}&=&\partial_{\mu}\psi_{\nu_e}-\frac i 2 g_{1}[\psi_{\nu_e},B_{\mu}]_\star,\nonumber \\
D_{\mu}^{3}\psi_d&=&\partial_{\mu}\psi_d-\frac{i}{2\sqrt{6}}g_{3}\psi_d\star G_{\mu}^{\circ}-\frac i 2 g_{3}\psi_d\star G_{\mu}^{a}T^{a},\nonumber \\
D_{\mu}^{1+3}\psi_u&=&\partial_{\mu}\psi_u+\frac i 2
g_{1}B_{\mu}\star\psi_u -\frac{i}{2\sqrt{6}}g_{3}\psi_u\star
G_{\mu}^{\circ}-\frac i 2 g_{3}\psi_u\star
G_{\mu}^{a}T^{a}.\label{7}\eea Moreover $U_\star(n)$ group can be
decomposed as follows \be U_\star(n)=U_n(1)\star NCSU(n), \ee where
$U_n(1)$ consists the Abelian $\th$-independent elements of
$U_\star(n)$ while $NCSU(n)$ consists the remaining parts.  In other
words the elements of $ U_\star(n)$ can be uniquely written as \be
U(x,\th)=e^{i\epsilon_0(x)1_n}\star e_{\star}^{i\epsilon_1(x,\th)1_n
+i\epsilon_a(x,\th)T^a}, \ee in which the first exponential factor
is a $\th$-independent function and by multiplying two different
elements of $U_\star(n)$ one can easily show that this factor group
is isomorphic to the usual commutative local gauge group $ U_n(1)$.
Therefore this one-dimensional representation of $U_\star(n)$ can be
considered to rewrite the gauge potential $A(x,\th)$ as

\be A(x,\th)=A_0(x)1_n+ A_1(x,\th)1_n+iA^0_a(x,\th)T^a,\ee where one
can show that under $U_\star(n)$-transformation, $A_0(x)$ transforms
as the usual U(1) gauge field.  In other words for the first step
symmetry breaking a commutative  scalar filed ( which was called
Higgsac ) is enough to reduce the extra $U(1)$-field.
 In fact in the $U_\star(3)\times U_\star(1)$ gauge theory there
exist two $U(1)$ factors  which can be reduced to one through an
appropriate Higgs mechanism. This extra scalar field is charged
under the both sub-groups, $U_{1}(1)$ and $U_{3}(1)$. The gauge
transformation of the Higgsac field for the first symmetry breaking
is \be\phi(x)\rightarrow
 U_{3}(x)\phi(x)V_1^{-1}(x),\label{8}\ee
where $U_{3}(x)\in U_3(1)$ and $V_1(x)\in U_{1}(1)$, also here
$\phi(x)$ is $\theta$-independent and the multiplication is in the
commutative space. Therefore the gauge invariant terms including the
Higgsac field are \be D_{\mu}^{1+1}\phi(x)D_{\mu}^{1+1}\phi(x) +
m^{2}\phi^{\dagger}(x)\phi(x) -
\frac{F}{4!}(\phi^{\dagger}(x)\phi(x))^{2},\label{9}\ee
 where
\be D_{\mu}^{1+1} = \partial_{\mu} +
i\frac{3g_{3}}{\sqrt{6}}G_{\mu}^{\circ} -
i\frac{g_{1}}{2}B_{\mu}.\label{10}\ee Here $B_{\mu}$ and
$G_{\mu}^{\circ}$ are the $\theta$-independent parts of their
corresponding  noncommutative fields. After the symmetry reduction
we shall obtain a massive gauge boson, ${G_{\mu}^{\circ}}^\prime$,
and a massless one, $A_\mu$. These mass eigenstates can be obtain in
terms of the gauge eigenstates as follows
%By performing a
%rotation in the $(B_{\mu},G_{\mu}^{\circ})$ plane, a massive field
%$(G_{\mu}^{\circ})$ and a massless field $(A_{\mu})$ are obtained
%which $A_{\mu},G_{\mu}^{\circ}$ fields are :
 \bea {G_{\mu}^{\circ}}^\prime&=&\cos\delta_{13}G_{\mu}^{\circ}-\sin\delta_{13}B_{\mu},\nonumber\\
A_{\mu}&=&\cos\delta_{13}B_{\mu}+\sin\delta_{13}G_{\mu}^{\circ},\label{11}\eea
in which the angle $\delta_{13}$ is
\be\delta_{13}=\tan^{-1}(\sqrt{\frac{2}{3}}\frac{g_{1}}{g_{3}}).\label{12}\ee
Meanwhile the mass of ${G_{\mu}^{\circ}}^\prime$can be obtained as
 \be M^{2} = \frac{1}{4}(g_{1}^{2} +\frac 3 2 g_{3}^{2})\phi^2_0,\label{13}\ee
where $\phi^2_0$ is the vacuum expectation value of $\phi$. Although
the $\th$-independent parts of $B_{\mu}$ and $G_{\mu}^{\circ}$ only
present in the Higgsac Lagrangian, we replace (\ref{11}) in the full
Lagrangian. However we have to notice that only the
$\th$-independent parts of ${G_{\mu}^{\circ}}^\prime$ gets mass
through Higgsac mechanism. Hereafter we use $G_{\mu}^{\circ}$
instead of ${G_{\mu}^{\circ}}^\prime$. In this manner, the fermions
of the $U_\star(3)\times U_\star(1)$ theory couple to the massless
gauge boson of the residual $U_\star(1)$, $A_\mu$, through the usual
electric charges if we define \be \frac 1 2 g_1\cos
\delta_{13}=e,\ee \be-\frac{1}{2\sqrt{6}} g_3\sin \delta_{13}=q_d ,
\ee where $q_d$ is the electric charge of the down quark, $-\frac 1
3 e$, and \be\frac 1 2 (g_1\cos
\delta_{13}-\frac{1}{\sqrt{6}}g_3\sin \delta_{13})=q_u, \ee where
$q_u$ is the electric charge of the up quark, $\frac 2 3 e$.
Therefore, the QED interactions of fermions in this model are as
 \be {\cal
L}_{e-A_\mu}=-ie\bar{\psi}_e\star\gamma^\mu\psi_e\star A_\mu,\ee \be
{\cal L}_{d-A_\mu}=-\frac i 3
e\bar{\psi}_d\star\gamma^\mu\psi_d\star A_\mu,\ee \be{\cal
L}_{u-A_\mu}=\frac{2i}{3}e\bar{\psi}_u\star\gamma^\mu
A_\mu\star\psi_u-\frac 1 3
e\bar{\psi}_u\star\gamma^\mu[\psi_u,A_\mu]_\star, \ee for electron,
down quark and up quark,  respectively. Finally, neutrinos which are
the neutral particles can be coupled to photons in the NC space-time
through the adjoint representation as \be {\cal
L}_{\nu-A_mu}=-ie\bar{\psi}_\nu\star\gamma^\mu[\psi_\nu,A_\mu]_\star.\ee
The Feynman rules for this model are completely collected in
Appendix B.
%########################################################################
\section{The Quantization of $U_{\star}(3)\times U_{\star}(1)$}
 The $U_\star(n)$ is a  non-Abelian gauge theory for all $n$, therefore
we should perform the gauge fixing to have the nonsingular propagator. The
Faddev-Popov and the gauge fixing terms for this theory are given as
\cite{hayakawa} \be S_{GF}=\int
d^dx(-\frac{1}{2\alpha}Tr(\partial_\mu G^\mu\partial_\nu
G^\nu)+\frac 1 2 Tr (i\bar{c}\star\partial_\mu D^\mu c-i\partial_\mu
D^\mu c\star\bar{c})), \ee where $c$ and $\bar{c}$ are the  ghost fields
and
 \be
 D_\mu c=\p_\mu c-ig[G_\mu,c]_\star,
 \ee
in which one needs $n^2$ ghost fields for the $n^2$ gauge fields.  Hence for
 the  $U_{\star}(3)\times U_{\star}(1)$ gauge group there
are nine ghosts for $U_{\star}(3)$ and one ghost for $U_{\star}(1)$ which can be written as
 \bea
 S_{GF}&=&\int d^dx(-\frac{1}{2\alpha}\partial_\mu B^\mu\partial_\nu B^\nu
+\frac 1 2 (i\bar{c}_B\star\partial_\mu D^\mu c_B-i\partial_\mu
D^\mu c_B\star\bar{c}_B))\no\\&+&\int
d^dx(-\frac{1}{\alpha}Tr(\partial_\mu G^\mu\partial_\nu G^\nu)+Tr
(i\bar{c}\star\partial_\mu D^\mu c-i\partial_\mu D^\mu
c\star\bar{c})),
 \eea
 where $c_B$ and $c$ are the ghost fields corresponding to $B_\mu$ and $G_\mu$, respectively, and
 \bea
 D_\mu c_B&=&\p_\mu c_B-ig_1[B_\mu,c_B]_\star,\no\\
 D_\mu c&=&\p_\mu c-ig_3[G_\mu,c]_\star.
 \eea
After the symmetry reduction, as was discussed in the pervious section,
the $\th$-independent part of $G_{\mu}^{\circ}$ gets mass through
the symmetry breaking while the $\theta$-dependent part remains
massless.
Therefore it is necessary to take into account a ghost
field corresponding to the $\theta$-dependent part of
$G_{\mu}^{\circ}$ which we denote by $c^\circ$. In the case of
$A_\mu(x)$ the $\th$-independent as well as $\th$-dependent part
remains massless. But the $\theta$-dependent part of $A_\mu(x)$ is
still non-Abelian and also needs a ghost field which we denote by
$c^\gamma$. These new ghost fields can be written in terms of $c_B$
and $c_{G^\circ}$ ( corresponding to $B_\mu$ and
$G^\circ_\mu$) as follows
   \bea
   c^{\circ}&=&\cos\delta_{13}c_{G^\circ}-\sin\delta_{13}c_B,\nonumber\\
c^\gamma&=&\cos\delta_{13}c_B+\sin\delta_{13}c_{G^\circ}.
   \eea
%=========================================================================
\section{The calculation of $\beta-$function}

A direct one loop calculation in the $U_{\star}(1)$ gauge theory for
both fundamental and adjoint representation have been resulted in a negative
$\beta-$function \cite{hayakawa,sadooghi}. The negative contribution
comes from the photon self interaction due to the NC space-time.
However, in the $U_\star(3)\times U_\star(1)$ gauge theory
the result will be definitely different.  In fact the reasons are two folds, the first one is that
the charge quantization problem is solved and quarks as well as the
 leptons are accommodated in the theory.  The second one is the appearance of the new interactions among the different gauge bosons i.e. photon and gluon, that makes the result complicated.
In this section we explicitly calculate the QED $\beta$-function in
the $U_\star(3)\times U_\star(1)$ theory. The QED $\beta$-function
can be obtained by the following relation \be \beta(e)=M\frac{\p}{\p
M}(-\delta_1+e\delta_2+\frac e 2 \delta_3),\ee in which $\delta_1$,
$\delta_2$ and $\delta_3$ are the vertex, fermion and photon
counter-terms, respectively.

%=========================================================================
\subsection{Electron self energy }

The one loop corrections for the electron self energy  are shown in
Fig.1 in which Fig.1(a) corresponds  to the ordinary QED. The UV
divergences of these diagrams can be subtracted by the rescaling of
the field and mass of the electron. The electron renormalization
factor $Z_{\psi}$ can be easily found as ( see appendix A,
Eq.(\ref{fermion}))
 \bea\delta
Z_{\psi}^{a} &=& -\frac{e^{2}}{16\pi^{2}}\frac{1}{\acute{\epsilon}},\nonumber\\
\delta Z_{\psi}^{b} &=& -\frac{e^{2}}{16\pi^{2}}\frac{1}{\acute{\epsilon}}(\tan^{2}\delta_{13}),\nonumber\\
 Z_{\psi} &=& 1 - \frac{e^{2}}{16\pi^{2}}\frac{1}{\acute{\epsilon}}(1 +
 \tan^{2}\delta_{13}),\label{14}\eea
where $\frac{1}{\acute{\epsilon}} =
 \frac{1}{\epsilon}+\gamma_{E}-\log4\pi$ for the space-time
 dimension $d = 4-2\epsilon $.  In presenting the explicit
 expressions of all renormalization constants the MMS scheme is used
 throughout.
\begin{figure}
\centerline{\epsfysize=1in\epsfxsize=4in\epsffile{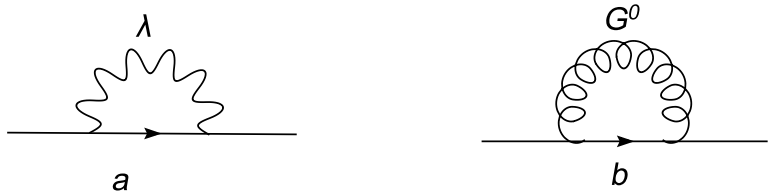}}\caption{
The electron self energy
 }
\label{fig_1}
\end{figure}

\subsection{Photon self energy }

The one loop contributions of the photon self energy are shown in
Fig.2.  The fermion loop in the  Fig.2(a) can be each one of
electron, neutrino, up and down quark.  The main difference in each
loop with respect to the ordinary space is an appearance of a phase
factor depends on the NC-parameter.  These factors for the charged
fermions fortunately cancel each other and therefore the
loop-calculations are completely like the ordinary $SU(3)\times
U(1)$. The contributions of such diagrams are ( see
Eq.(\ref{chargedfermion})):
\begin{eqnarray}
\delta Z_{A}^{cf} &=&
-\frac{e^{2}}{16\pi^{2}}\frac{1}{\acute{\epsilon}}\frac{4}{3}(1 +3(
(2/9 + 4/9 + 4/9) + 1/9)),\nonumber\\
&=&
-\frac{e^{2}}{16\pi^{2}}\frac{1}{\acute{\epsilon}}\frac{4}{3}(\frac{14}{3})\label{15},
\end{eqnarray} where $cf$ means charged fermion.
\begin{figure}
\centerline{\epsfysize=3.5in\epsfxsize=6in\epsffile{
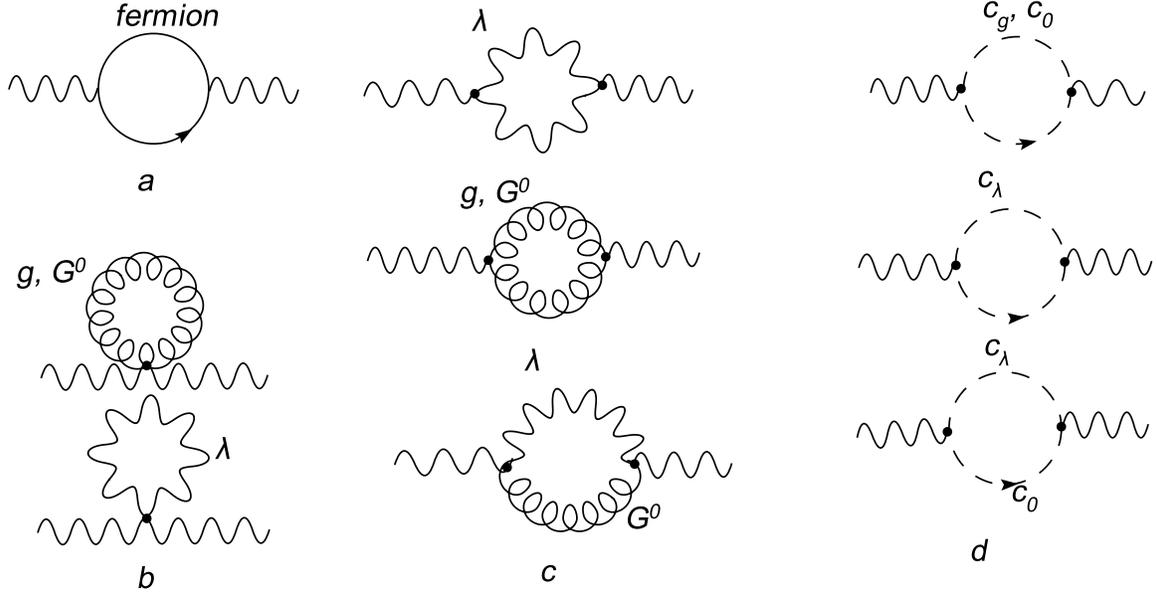}}\caption{Photon self energy  (a) fermion loop (b,c) gauge
boson loop (d)ghost boson loop
 }
\label{figA_2}
\end{figure}

Meanwhile the neutrino loop contribution in the photon self energy
is different from the charged one, because the noncommutative phases
in this case do not cancel each other. Therefore for such a diagram
one has ( see Eq.(\ref{neutrino})) \be\delta Z_{A}^{nf} =
-\frac{e^{2}}{16\pi^{2}}\frac{1}{\acute{\epsilon}}\frac{4}{3}(2),\label{16}\ee
where $nf$ means neutral fermion. Then  the total contributions from
the fermion loops i.e. charged fermions (lepton and quark) and
neutrino, are
 \begin{eqnarray}\delta Z_{A}^{f} &=&
-\frac{e^{2}}{16\pi^{2}}\frac{1}{\acute{\epsilon}}\frac{4}{3}N_{F}(1+33/9
+ 2),\nonumber\\
&=&
-\frac{e^{2}}{16\pi^{2}}\frac{1}{\acute{\epsilon}}\frac{4}{3}N_{F}(20/3
),\label{17}\end{eqnarray}
 where $N_{F}$ is the number of generations.\\
The remaining parts of the Fig.2 show  the  gauge boson one-loop
corrections to the photon self energy. The gauge boson diagrams in
Fig.2(b) have  singularity proportional to $\frac{1}{q^{2}}$ that
can be canceled by those come from the diagrams given in the
Fig.2(c). In fact these diagrams have not any contribution on the
$\beta-$function of $U_\star(3)\times U_\star(1)$ model
\cite{hayakawa}. Hence, the one loop corrections for the photon self
energy which contain the photon-gluon interactions induced by the NC
space-time can be obtained as ( see Eq.(\ref{gauge loop}) )

\begin{eqnarray}
% \nonumber to remove numbering (before each equation
\delta Z_{A}^{g} =
+\frac{e^{2}}{16\pi^{2}}\frac{1}{\acute{\epsilon}}\frac{10}{3}(1/9)(10+\delta_{ab})=
  +\frac{e^{2}}{16\pi^{2}}\frac{1}{\acute{\epsilon}}\frac{10}{3}(2).
\end{eqnarray}
 Therefore the total renormalization constant for the one-loop correction of the photon self energy is \be\delta Z_{A} =
-\frac{e^{2}}{16\pi^{2}}\frac{1}{\acute{\epsilon}}\frac{20}{3}[\frac{4}{3}N_{F}
-1].\label{19}\ee
\subsection{Vertex function }

After studying the UV divergences of two point function in the
$U_\star(3)\times U_\star(1)$ theory, now we have to calculate the
photon-electron vertex function  at the one-loop level. There are
four diagrams in this case as shown in the  Fig.3.  The three
diagrams in the Fig.3(b) show the NCQED effects in the
$U_\star(3)\times U_\star(1)$ theory.  It should be noted that here
there is two additional diagrams with respect to the NCQED based on
the  $ U_\star(1)$.  The $QED$ like diagram given in Fig.3(a) is
finite unlike the ordinary $QED$ \cite{hayakawa} but for the
nonabelian diagrams one can easily find ( see Eq.(\ref{vertex}))
%%%%%%%%%%%%%%%%%%%%%%%%%%%%%%%%%%%%%%%%
\begin{figure}
\centerline{\epsfysize=1in\epsfxsize=4in\epsffile{
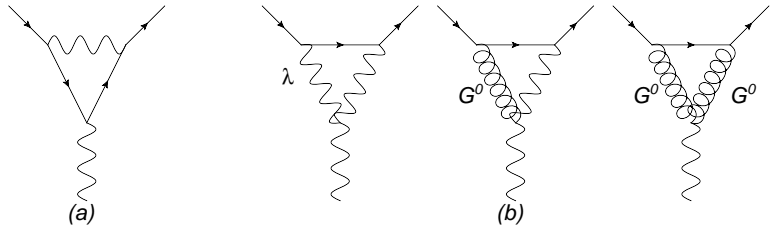}}\caption{Vertex function (a) $QED$ like diagram (b)These
diagrams occur through a non abelian vertex
 }
\label{figA_1}
\end{figure}
%%%%%%%%%%%%%%%%%%%%%%%%%%%%%%%%%%%%%%%%%%
\begin{figure}
\centerline{\epsfysize=3in\epsfxsize=3.5in\epsffile{
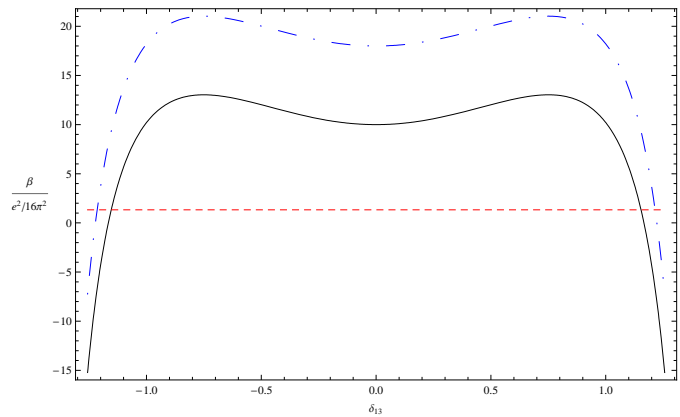}}\caption{$\frac{16\pi}{e^3}\beta$ versus $\delta_{13}$. Black
(solid), blue (dot) and red (dashed) curves represent the
$\beta$-function corresponding to NCQED without neutrinos, NCQED
with neutrinos and usual QED, respectively.} \label{beta}
\end{figure}

%%%%%%%%%%%%%%%%%%%%%%%%%%%%%%%%%%%%%%%%%%

\be\delta Z_{\bar{\psi}\psi A} =
-\frac{e^{2}}{16\pi^{2}}\frac{1}{\acute{\epsilon}}3\frac{1}{3}(1 +
 2\cos^{2}\delta_{13} + \tan^{2}\delta_{13}(1 + 2\sin^{2}\delta_{13}) - 2\tan\delta_{13}\sin2\delta_{13}).\label{20}\ee
 Now  the $\beta-$function of the NCQED part of the $U_\star(3)\times U_\star(1)$ gauge theory, by using (\ref{14}),(\ref{19}) and (\ref{20}), can be obtained as follows
 \bea\beta(e) =
+\frac{e^{3}}{16\pi^{2}}\{&+&\frac{4}{3}N_{F}(1+2
+ 33/9)+2(1 + \tan^{2}\delta_{13})\nonumber \\
&-&2[\frac{5}{3}(2) +
(1 + \cos^{2}\delta_{13})\nonumber \\
&+& \tan^{2}\delta_{13}(1 + 2\sin^{2}\delta_{13}) -2
\tan\delta_{13}\sin2\delta_{13}]\},\label{21}\eea where in the first
term the contributions of the charged leptons, neutrinos and quarks
are given separately.  One should note that the neutrino has only
contribution to the first term.  Meanwhile (\ref{21}) shows that the
$\beta$-function depends on the mixing parameter, $\delta_{13}$ as a
free parameter.  In fact the $\beta$, depending on the value of
$\delta_{13}$ can be negative, positive or even zero as shown in
Fig.4.  In the Fig.4 the $\beta$-function for the ordinary QED is
compared with the NCQED (with and without neutrino ).  It can be
easily seen that the NC QED $\beta$-function is equal to the usual
QED for $\delta_{13}=0.367\pi$.
%===========================================================================
\section{Conclusion}

We have considered the  unexpanded NCQED based on the $U_\star(3)\times U_\star(1)$ gauge group.
We have shown that the $\beta$ function in agreement with \cite{hayakawa} does not depend on the
 parameter of non commutativity but in contrast with the $ U_\star(1)$  NCQED it depends on a
 new parameter i.e. mixing parameter $\delta_{13}$, see (\ref{21}).   Therefore the $\beta$ function
 can be negative, positive or zero depending on the value of the mixing parameter. Meanwhile the $\beta$ function
 in the $ U_\star(1)$  NCQED  is suffering from the asymptotic freedom that is in contrast with the experiment.
 We have compared our result with the ordinary QED in the Fig.4 and it is found that for the $\delta_{13}=0.367\pi$ our $\beta$
 function is equal to the ordinary QED.  In fact we have shown that in contrast with the expanded QED the unexpanded NCQED based
 on the $U_\star(3)\times U_\star(1)$ gauge group is not only one-loop renormalizable but also by fixing $\delta_{13}$ it
 can have the same value for the $\beta$
 function as the commutative QED.  Nevertheless the obtained value for $\delta_{13}$  needs to confirm in an independent way.
 %%%%%%%%%%%%%%%%%%%%%%%%%%%%%%%%%%%%%%%%%%%%%%%%%%%%%%%%%%%%%%%%%%%%%%%%%%%%%%%%%%%%%%%%%%%%%%%%%%%%%%%%%%%%%%%%%%%%%%%%%%%%%%%%%

\appendix
\section{The calculations of two and three point functions} {\bf
Electron self energy:} The electron self-energy includes two
diagrams as are shown in the Fig.\ref{fig_1}. Using the Feynman
rules (\ref{e-p}) and (\ref{e-g0}) one can write

\begin{eqnarray}
  -i\Sigma(p) &=& (-ie)^2(1+\tan^2\delta_{13})\int\frac{d^4k}{(2\pi)^2}\gamma^{\mu}\frac{k\!\!\!/+m}{k^2-m^2}\gamma_{\mu} \frac{1}{(k-p)^2},\nonumber\\
  &=&
  (-ie)^2(1+\tan^2\delta_{13})\int_0^1dx\int\frac{d^dl}{(2\pi)^2}\frac{-2xp\!\!\!/+4m}{(l^2-\Delta)^2},
  \nonumber\\
\end{eqnarray}
where $\Delta=(1-x)m^2-x(1-x)p^2$.  Then one has
\begin{eqnarray}\label{fermion}
  \delta Z_\psi&=& \frac{d\Sigma}{d p\!\!\!/}|_{p\!\!\!/=m}, \nonumber\\
   &=&
   -\frac{e^2}{16\pi^2}(1+\tan^2\delta_{13})\frac{1}{\acute{\epsilon}}.
\end{eqnarray}
{\bf Photon self energy:} All diagrams corresponding to the one loop
photon self energy are shown in the Fig.2. First we calculate the
fermion loop, Fig.2a, containing electron, $u$ and $d$ quarks as
follows

\begin{equation}\label{}
    i\Pi_{cf}^{\mu\nu} = -(-ie)^2\int\frac{d^4k}{(2\pi)^2}(I_e+3I_u+3I_d)\frac{Tr\{\gamma^\mu(k\!\!\!/+m)\gamma^\nu(k\!\!\!/-\q
    +m)\}}{(k^2-m^2)((k-q)^2-m^2)},
\end{equation}
where the coefficient $3$ is a color factor for the quarks $u$ and
$d$.  Using the Feynman rules (\ref{e-p}), (\ref{d-p}) and
(\ref{u-p}) one can easily find
\begin{eqnarray}
  I_e &=& 1, \nonumber\\
  I_d &=& 1/9, \nonumber\\
  I_u &=& 4/9+2/9+8/9\sin^2(q\times k)=4/9+2/9+4/9+{\cal NP},
\end{eqnarray}
where ${\cal NP}$ means non-planer. In fact the non-planer part has
not any contribution to the beta function \cite{hayakawa} and it is
not needed to calculate this part. Meanwhile the planer parts are
similar to the ordinary QED and it can be easily obtained as
\begin{eqnarray}\label{chargedfermion}
   i\Pi_{cf}^{\mu\nu}&=&e^2(1+33/9)\int_0^1dx\int\frac{d^dl}{(2\pi)^2}\frac{g^{\mu\nu}(2/d-1)l^2-2x(1-x)q^\mu q\nu+g^{\mu\nu}(m^2+x(1-x)q^2)}{(l^2-\Delta)^2},\nonumber\\
   &=&-ie^2(1+33/9)(q^2g^{\mu\nu}-q^\mu
   q\nu)\frac{8}{16\pi^2}\int^1_0x(1-x)\frac{1}{\acute{\epsilon}}+{\cal NP},\nonumber\\
   &=&-ie^2(1+33/9)(q^2g^{\mu\nu}-q^\mu
   q^\nu)\frac{1}{16\pi^2}\frac{4}{3}\frac{1}{\acute{\epsilon}}+{\cal
   NP},
\end{eqnarray}
where $\Delta=m^2-x(1-x)q^2$.  Similarly, the impact of the neutrino
loop on the photon self energy, using (\ref{n-p}), can be easily
obtained as follows
\begin{eqnarray}\label{neutrino}
  i\Pi_{neutrino}^{\mu\nu} &=& -(-ie)^2\int\frac{d^4k}{(2\pi)^2}(4\sin^2(q\times k))\frac{Tr\{\gamma^\mu k\!\!\!/ \gamma^\nu (k\!\!\!/-\q) \}}{(k^2)((k-q)^2)},\nonumber \\
  &=& 2e^2\int_0^1dx\int\frac{d^dl}{(2\pi)^2}\frac{g^{\mu\nu}(2/d-1)l^2-2x(1-x)q^\mu
  q_\nu+x(1-x)q^2g^{\mu\nu}}{(l^2-\acute{\Delta})^2}+{\cal NP},\nonumber\\
  &=&-2ie^2(q^2g^{\mu\nu}-q^\mu
  q^\nu)\frac{1}{16\pi^2}\frac{4}{3}\frac{1}{\acute{\epsilon}}+{\cal
  NP},
\end{eqnarray}
where $\acute{\Delta}=-x(1-x)q^2$. Therefore the total contribution
of the fermion loops can be obtained as
\begin{equation}\label{fermion loop}
  \delta Z^{f}_A=-\frac{e^2}{16\pi^2}\frac{4}{3}N_F(1+2+33/9).
\end{equation}
Now we consider the contribution of the gauge bosons on the
self-energy of photon, Fig.2b and Fig.2c. Let us start with the
diagrams given in Fig.2b. Using the Feynman rules (\ref{4p}) and
(\ref{2p-2g0}), the corresponding self energy can be found as
follows
\begin{eqnarray}
  i\Pi_{Fig.2b}^{\mu\nu} &=& -\frac{1}{2}e^2(2/3)^2(10+\delta_{ab})\int\frac{d^4k}{(2\pi)^2}2(d-1)g^{\mu\nu}\sin^2(q\times k)\frac{1}{k^2}, \nonumber\\
  &=&-\frac{1}{2}e^2(2/3)^2(10+\delta_{ab})\int\frac{d^4k}{(2\pi)^2}(d-1)g^{\mu\nu}\frac{1}{k^2}\frac{(q+p)^2}{(q+p)^2}+{\cal
  NP},
\end{eqnarray}
or for the planar part
\be
 i\Pi_{Fig.2b}^{\mu\nu}=-e^2\frac{1}{2}(2/3)^2(10+\delta_{ab})\int_0^1dx\int\frac{d^dl}{(2\pi)^2}\frac{(d-1)g^{\mu\nu}(l^2+x(1-x)q^2)}{(l^2-\Delta)^2},
\ee
 where $\Delta=-x(1-x)p^2$, $1/2$ is a symmetry factor and the dimensional
regularization is used throughout. Then after performing the
integral on the momenta we have
\begin{eqnarray}
  i\Pi_{Fig.2b}^{\mu\nu} &=& \frac{ie^2}{16\pi^2}\frac{1}{2}(2/3)^2(10+\delta_{ab})\int_0^1dx\frac{q^2g^{\mu\nu}}{\Delta^{2-d/2}}\nonumber \\
   &(&-\Gamma(1-d/2)d/2(d-1)x(1-x)-\Gamma(2-d/2)(d-1)(1-x)^2).
\end{eqnarray}
The contribution of the diagrams given in the Fig.2c on the photon
self energy can be easily obtained in a similar way as follows
\begin{eqnarray}
  i\Pi_{Fig.2c}^{\mu\nu} &=&  -\frac{1}{2}e^2(2/3)^2(10+\delta_{ab})\int\frac{d^4k}{(2\pi)^2}\frac{1}{k^2}\frac{1}{(q+k)^2} \nonumber\\
  && [-g^{\mu\nu}[(2k+q)^2+(q-p)^2]-d(k+2q)^\mu(k+2q)^\nu +[(2k+q)^\mu(k+2q)^\nu\nonumber\\
  &&+
  (k-q)^\mu(2k+q)^\nu-(k+2q)^\mu(k-q)^\nu+(\mu\leftrightarrow\nu)]],
\end{eqnarray}
or
\begin{eqnarray}
  i\Pi_{Fig.2c}^{\mu\nu} &=&  -\frac{1}{2}e^2(2/3)^2(10+\delta_{ab})\int_0^1dx\int\frac{d^dl}{(2\pi)^2} \frac{1}{(l^2-\Delta)^2}[-g^{\mu\nu}(l^26(1-1/d)\nonumber\\
  &&-q^2(3-2x+2x^2)) +q^\mu q^\nu[(2-d)(1-2x)^2+2(1+x)(2-x)]],
\end{eqnarray}
and after performing the integral on the momentum the self energy
leads to
\begin{eqnarray}
  i\Pi_{Fig.2c}^{\mu\nu} &=& \frac{ie^2}{16\pi^2}\frac{1}{2}(2/3)^2(10+\delta_{ab})\int_0^1dx\frac{q^2g^{\mu\nu}}{\Delta^{2-d/2}}\nonumber\\
   &&[\Gamma(1-d/2)g^{\mu\nu}q^2[3/2(d-1)(x-x^2)]+\Gamma(2-d/2)g^{\mu\nu}q^2(1/2)[3-2x+2x^2]\nonumber\\
   &&-\Gamma(2-d/2)q^\mu q^\nu(1/2)[(2-d)(1-2x)^2+2(1+x)(2-x)]].
\end{eqnarray}
Here, one should note that for the two photons (using (\ref{3p})),
two $\acute{G}^0$ (using (\ref{p-2g0})) and two gluons (\ref{p-2g})
in the loop the symmetry factor is $1/2$ but for the
photon-$\acute{G}^0$ (using (\ref{g0-2p})) in the loop the factor is
one. Finally, we must include the diagrams containing ghost
loops(Fig.2d). The corresponding Feynman rules are photon-ghost of
photon (\ref{p-gh(p)}); photon-ghost of $G^\circ_\mu$
(\ref{p-gh(g0)}); photon-ghost of gluons (\ref{p-gh(g)});
photon-ghost of photon and $G^\circ_\mu$ (\ref{p-gh(p-g0)}).
Therefore for the corresponding  self energy one has

\begin{equation}\label{}
     i\Pi_{Fig.2d}^{\mu\nu} = -e^2(2/3)^2(10+\delta_{ab})\int\frac{d^4k}{(2\pi)^2}\sin^2(q\times
     k)\frac{i(k+q)^\mu}{k^2}\frac{ik^\nu}{(q+k)^2},
\end{equation}
or after manipulating some algebra one finds
\begin{eqnarray}
  i\Pi_{Fig.2d}^{\mu\nu} &=& \frac{ie^2}{16\pi^2}\frac{1}{2}(2/3)^2(10+\delta_{ab})\int_0^1dx \frac{q^2g^{\mu\nu}}{\Delta^{2-d/2}}\nonumber\\
  &(&-\Gamma(1-d/2)g^{\mu\nu}q^2[x(1-x)/2]+\Gamma(2-d/2)q^\mu
  q^\nu[x(1-x)]),
\end{eqnarray}
which leads to
\begin{eqnarray}\label{gauge loop}
  i\Pi_{gauge}^{\mu\nu} &=& i\Pi_{Fig.2d}^{\mu\nu}+i\Pi_{Fig.2c}^{\mu\nu}+i\Pi_{Fig.2b}^{\mu\nu}, \nonumber\\
  &=& i(q^2g^{\mu\nu}-q^\mu
  q^\nu)\frac{10}{3}(1/9)(10+\delta_{ab})\frac{e^2}{16\pi^2}\frac{1}{\acute{\epsilon}}+...,
\end{eqnarray}
and therefore
\begin{equation}\label{}
     \delta Z_A =
     -\frac{e^2}{16\pi^2}\frac{1}{\acute{\epsilon}}[\frac{4}{3}N_F(1+2+33/9)-\frac{10}{3}(1/9)(10+\delta_{ab})].
\end{equation}

Now we consider the vertex function and the corresponding diagrams
which are given in the Fig.3. The diagram given in the Fig.3a is
very similar to its counterpart in the ordinary $QED$ but with
different divergence behavior.  Here using the NC rules one can
easily find
\begin{eqnarray}
  -ie\Gamma^\mu &=& -ie \exp(i p\times \acute{p})(\gamma^\mu+\delta\Gamma^\mu),\nonumber \\
  \delta\Gamma_{Fig.3a}^\mu&=& -ie^2(1+\tan^2\delta_{13})\int\frac{d^4k}{(2\pi)^4}\frac{\exp(i q\times k)\gamma^\lambda(k\!\!\!/+\q+m)\gamma^\mu(k\!\!\!/+m)\gamma_\lambda}
  {(k^2-m^2)((k+q)^2-m^2)(p-k)^2},
\end{eqnarray}
where the factor $\exp(i q\times k)$ causes the equation to be
finite. Meanwhile for the diagrams given in Fig.3b we have
\begin{eqnarray}
  \delta\Gamma_{Fig.3b}^\mu &=&+(2/3)e^2I_v \int\frac{d^4k}{(2\pi)^4}\frac{2i\sin^2(q\times k)+2\cos(q\times k)\sin(q\times k)}{k^2(k+q)^2((p-k)^2-m^2)}\nonumber \\
  &\times&(\gamma^\lambda(\p-k\!\!\!/+m)\gamma^\rho)[-(2k+q)^\mu g^{\rho\lambda}+(2q+k)^\rho g^{\mu\lambda}+(k-q)^\lambda
  g^{\mu\rho}],
\end{eqnarray}
or
\begin{equation}\label{}
    \delta\Gamma_{Fig.3b}^\mu =+i(2/3)e^2I_v \int\frac{d^4k}{(2\pi)^4}\frac{-4k\!\!\!/
    k^\mu-2k^2\gamma^\mu}{k^2(k+q)^2((p-k)^2-m^2)}+ ...,
\end{equation}
 where
\begin{equation}\label{}
    I_v=1+2\cos^2\delta_{13}
    +(1+2\sin^\delta_{13})\tan^2\delta_{13}-2\tan\delta_{13}(\sin2\delta_{13}),
\end{equation}
and $...$ means the finite part of the integral.  Therefore $\delta
Z_{\bar{\psi}A\!\!\!/\psi}$ can be easily found as follows

\begin{equation}\label{vertex}
    \delta Z_{\bar{\psi}A\!\!\!/\psi}=-\frac{e^2}{16\pi^2}3(1/3)(1+2\cos^2\delta_{13}
    +(1+2\sin^\delta_{13})\tan^2\delta_{13}-2\tan\delta_{13}(\sin2\delta_{13}))\frac{1}{\acute{\epsilon}}.
\end{equation}

%%%%%%%%%%%%%%%%%%%%%%%%%%%%%%%%%%%%%%%%%%%%%%%%%%%%%%%%%%%%%%%%%%%%%%%%%%%%%%%%%%%%%%%%%%%%%%%%%%%%%%%%%%%%%%%%%%%%%%%%%

%=========================================================================
%\appendix
\section{Feynman rules for  $NC(SU(3)\times U(1))$ model.} Besides
the fermion part of the action that is given in section 2, after
symmetry reduction, one can obtain the action of the gauge
interactions in terms of the mass eigenstates that is given in the
first part of this appendix.  In the second part we give the Feynman
rules for  $NC(SU(3)\times U(1))$ model.  Here the fields are
physical and for simplicity $G_{\mu}^{\circ}$ is used instead of
$G^{0'}_{\mu}$ as follows
 \bea
S_{photon}=-\frac 1 4 \int
\{\p_{[\mu}A_{\nu]}\star\p^{[\mu}A^{\nu]}+\frac 4 3 ie(1+
2\cos^2\delta)\p_{[\mu}A_{\nu]}\star
[A^\mu,A^\nu]_\star\nonumber\\
-\frac 4 9
e^2(1+8\cos^2\delta_{13})[A_\mu,A_\nu]_\star\star[A^\mu,A^\nu]_\star\}
d^4x,\eea
%%%%%%%%%%%%%%%%%%%%%%%%%%%%%%%%%%%%%%%%%%%%%%%%%%%%%%%%%%%%%%%%%%%%%%%%
\bea S_{G_{\mu}^{\circ}}=-\frac 1 4 \int
\{&\hspace{-3mm}\p_{[\mu}G^{\circ}_{\nu]}\star\p^{[\mu}{G^{\circ}}^{\nu]}-2ie(2\tan\delta_{13}\sin^2\delta_{13}-\frac
2 3 \cot\delta_{13}\cos^2\delta_{13})\p_{[\mu}G^{\circ}_{\nu]}\star
[{G^{\circ}}^\mu,{G^{\circ}}^\nu]_\star\nonumber\\
&\hspace{-10mm}-4e^2(\tan^2\delta_{13}\sin^2\delta_{13}+
\frac{1}{9}\cot^2\delta_{13}\cos^2\delta_{13})[{G^{\circ}}_\mu,G^{\circ}_\nu]_\star\star[{G^\circ}^\mu,{G^\circ}^\nu]_\star\}
d^4x,\eea
%%%%%%%%%%%%%%%%%%%%%%%%%%%%%%%%%%%%%%%%%%%%%%%%%%%%%%%%%%%%%%%%%%%%%%%%%%
\bea S_{photon-G_{\mu}^{\circ}}=-\frac 1 4
\int\{&&\hspace{-7mm}-\frac 4 3
ie\sin2\delta_{13}\Big(\p_{[\mu}A_{\nu]}\star([A^\mu,{G^\circ}^\nu]_\star+[{G^\circ}^\mu,A^\nu]_\star)\no\\
&&\hspace{-10mm}+\p_{[\mu}G^{\circ}_{\nu]}\star[A^\mu,A^\nu]_\star\Big)\no\\&&\hspace{-10mm}+4ie(1-\frac
2 3
\cos^2\delta_{13})\Big(\p_{[\mu}G^{\circ}_{\nu]}\star([A^\mu,{G^\circ}^\nu]_\star+[{G^\circ}^\mu,A^\nu]_\star)\no\\&&\hspace{-10mm}
+\p_{[\mu}A_{\nu]}\star[{G^\circ}^\mu,{G^\circ}^\nu]_\star\Big)\no\\&&\hspace{-10mm}+
\frac{16}{9}e^2\sin2\delta[A_\mu,A_\nu]_\star\star([A^\mu,{G^\circ}^\nu]_\star+[{G^\circ}^\mu,A^\nu]_\star)\no\\&&\hspace{-10mm}+
4e^2(\tan\delta_{13}\sin^2\delta_{13}-\cot\delta_{13}\cos^2\delta_{13})[{G^\circ}^\mu,{G^\circ}^\nu]_\star\star
([{G^\circ}^\mu,A^\nu]_\star+[A^\mu,{G^\circ}^\nu])\no\\&&\hspace{-10mm}
-8e^2(1-\frac 8 9
\cos^2\delta_{13})[A_\mu,A_\nu]_\star\star[{G^\circ}^\mu,{G^\circ}^\nu]_\star\}d^4x,
\eea
%%%%%%%%%%%%%%%%%%%%%%%%%%%%%%%%%%%%%%%%%%%%%%%%%%%%%%%%%%%%%%%%
\bea S_{gluon}=-\frac 1 4
\int\{&&\hspace{-7mm}\p_{[\mu}G^{a}_{\nu]}\star\p^{[\mu}G_{a}^{\nu]}+2i\sqrt{\frac
2 3 }\frac{e}{\sin\delta_{13}}(
if_{abc}\p_{[\mu}G^{c}_{\nu]}\star\{{G^a}^\mu,{G^b}^\nu\}_\star\no\\&&\hspace{-10mm}
+d_{abc}\p_{[\mu}G^{c}_{\nu]}\star[{G^a}^\mu,{G^b}^\nu]_\star)\no\\&&\hspace{-10mm}-\frac
6 9
\frac{e^2}{\sin^2\delta_{13}}\Big(-f_{abc}f^{dec}\{G^a_\mu,G^b_\nu\}_\star\star\{G_d^\mu,G_e^\nu\}_\star\\&&\hspace{-10mm}
+2if_{abc}d^{dec}\{G^a_\mu,G^b_\nu\}_\star\star[G_d^\mu,G_e^\nu]_\star+d_{abC}d^{deC}[G^a_\mu,G^b_\nu]_\star\star[G_d^\mu,G_e^\nu]_\star\Big)\}d^4x,\no
\eea
%%%%%%%%%%%%%%%%%%%%%%%%%%%%%%%%%%%%%%%%%%%%%%%%%%%%%%%%%%%%
\bea S_{gluon-photon}=-\frac 1 4 \int\{&&\hspace{-7mm}2i\sqrt{\frac
2 3
}ed_{ab0}\p_{[\mu}A_{\nu]}\star[G^a_\mu,G^b_\nu]_\star+4ie\sqrt{\frac
2 3
}\p_{[\mu}G^{b}_{\nu]}\star([G^\mu_b,A^\nu]_\star+[A^\mu,G^\nu_b]_\star)\no\\&&\hspace{-10mm}-\frac
8 9
e^2\delta_{ab}[A_\mu,A_\nu]_\star\star[{G^a}^\mu,{G^b}^\nu]_\star\\&&\hspace{-10mm}
-\frac{8e^2}{3\sin\delta_{13}}(if_{abc}\{G^a_\mu,G^b_\nu\}_\star+d_{abc}[G^a_\mu,G^b_\nu]_\star)\star([{G^c}^\mu,A^\nu]_\star+[A^\mu,{G^c}^\nu]_\star)\}
d^4x,\no \eea
%%%%%%%%%%%%%%%%%%%%%%%%%%%%%%%%%%%%%%%%%%%%%%%%%%%%%%%%%%%%%%%%%%%%%%%%%%%
\bea S_{gluon-G^\circ_\mu}=-\frac 1 4
\int\{&&\hspace{-7mm}2i\sqrt{\frac 2 3
}e\cot\delta_{13}d_{ab0}\p_{[\mu}G^\circ_{\nu]}\star[G^a_\mu,G^b_\nu]_\star\no\\&&\hspace{-10mm}
+4ie\sqrt{\frac 2 3
}\cot\delta_{13}\p_{[\mu}G^{b}_{\nu]}\star([G^\mu_b,{G^\circ}^\nu]_\star+[{G^\circ}^\mu,G^\nu_b]_\star)\no\\&&\hspace{-10mm}
-\frac 8 9
e^2\cot^2\delta_{13}\delta_{ab}[G^\circ_\mu,G^\circ_\nu]_\star\star[{G^a}^\mu,{G^b}^\nu]_\star\no\\&&\hspace{-10mm}
-\frac{8e^2\cot\delta_{13}}{3\sin\delta_{13}}(if_{abc}\{G^a_\mu,G^b_\nu\}_\star+d_{abc}[G^a_\mu,G^b_\nu]_\star)\star([{G^\circ}^\mu,{G^c}^\nu]_\star\no\\
+[{G^c}^\mu,{G^\circ}^\nu]_\star)\}d^4x, \eea
%%%%%%%%%%%%%%%%%%%%%%%%%%%%%%%%%%%%%%%%%%%%%%%%%%%%%%%%%%%%%%%%%%%%%%%%%
\bea S_{gluon-G^\circ_\mu-photon}=-\frac 1 4 \int\{-\frac 8 9
e^2\cot\delta_{13}\delta_{ab}[G^a_\mu,G^b_\nu]_\star\star([{A}^\mu,{G^\circ}^\nu]_\star+[{G^\circ}^\mu,{A}^\nu]_\star)\}d^4x,\no\\
\eea
%%%%%%%%%%%%%%%%%%%%%%%%%%%%%%%%%%%%%%%%%%%%%%%%%%%%%%%%%%%%%%%%%%%%%%%%%%%%%%%%%%%%%%%%%%%%%%%%%%
\bea S_{ghosts-gauges}=\int \{&&\hspace{-7mm}e(1-\frac 2 3
\sin^2\delta_{13})\bar{c}^\gamma\star\p^\mu[A_\mu,c^\gamma]_\star-\frac
e 3
\sin2\delta_{13}\bar{c}^\gamma\star\p^\mu[G^\circ_\mu,c^\gamma]_\star\no\\&&\hspace{-10mm}
+(-e\tan\delta_{13}\sin^2\delta_{13}+\frac e 3
\cot\delta_{13}\cos^2\delta_{13})\bar{c}^\circ\star\p^\mu[{G^\circ}_\mu,c^\circ]_\star\no\\&&\hspace{-10mm}
+e(1-\frac 2 3
\cos^2\delta_{13})\big(\bar{c}^\circ\star\p^\mu[{A}_\mu,c^\circ]_\star+\bar{c}^\circ\star\p^\mu[{G^\circ}_\mu,c^\gamma]_\star
+\bar{c}^\gamma\star\p^\mu[{G^\circ}_\mu,c^\circ]_\star\big)\no\\&&\hspace{-10mm}
-\frac e 3
\sin2\delta\big(\bar{c}^\circ\star\p^\mu[{A}_\mu,c^\gamma]_\star+\bar{c}^\gamma\star\p^\mu[{A}_\mu,c^\circ]_\star\big)
+\frac e 3
\cot\delta_{13}\bar{c}_b\star\p^\mu[{G^\circ}_\mu,c^b]_\star
\no\\&&\hspace{-10mm}+\frac e 3
\bar{c}_b\star\p^\mu[{A}_\mu,c^b]_\star+\frac e 3
\bar{c}_b\star\p^\mu[{G^b}_\mu,c^\gamma]_\star+\frac e 3
\bar{c}^\gamma\star\p^\mu[{G^b}_\mu,c_b]_\star\no\\&&\hspace{-10mm}+\frac
e 3
\cot\delta_{13}\bar{c}_b\star\p^\mu[{G^b}_\mu,c^\circ]_\star+\frac e
3
\cot\delta_{13}\bar{c}^\circ\star\p^\mu[{G^b}_\mu,c_b]_\star\no\\&&\hspace{-10mm}
+ig_3f_{abc}\bar{c}^c\star\p^\mu\{{G^a}_\mu,c^b\}_\star
+g_3d_{abc}\bar{c}^c\star\p^\mu[{G^a}_\mu,c^b]_\star\}d^4x.\no\\
\eea
%%%%%%%%%%%%%%%%%%%%%%%%%%%%%%%%%%%%%%%%%%%%%%%%%%%%%%%%%%%%%%%%%%%%%%%%%%%%%%%%%%%%%%%%%%%%%%%%%%%
 Gluons carry Lorentz indices $\mu,\nu,...$, color indices $a,b,...$, and
momenta $p,q,...$. Ghosts carry only the last two type of labels.
All the momenta are entering unless otherwise specified.
%%%%%%%%%%%%%%%%%%%%%%%%%%%%%%%%%%%%%%%%%%%%%%%%%%%%%%%%%%%%%%%%%%%%%%%%%
\vskip .4cm\noindent{\bf electron--photon vertex}: \be \label{e-p}
-ie\exp{(ip\times k)}\gamma^\mu,\ee

%%%%%%%%%%%%%%%%%%%%%%%%%%%%%%%%%%%%%%%%%%%%%%%%%%%%%%%%%%%%%%%%%%%%%%%%%%%%%%%%%%%%
\vskip .4cm\noindent{\bf down quark--photon vertex}: \be
\label{d-p}-\frac 1 3 ie\exp{(ip\times k)}\gamma^\mu,\ee

%%%%%%%%%%%%%%%%%%%%%%%%%%%%%%%%%%%%%%%%%%%%%%%%%%%%%%%%%%%%%%%%%%%%%%%%%%%%%%%%%%%
\vskip .4cm\noindent{\bf up quark--photon vertex}: \be
\label{u-p}\frac 2 3 ie(\cos{(ip\times k)+2i\sin(p\times
q)})\gamma^\mu,\ee

%%%%%%%%%%%%%%%%%%%%%%%%%%%%%%%%%%%%%%%%%%%%%%%%%%%%%%%%%%%%%%%%%%%%%%%%%%%%%%%%%%%%%%%%%%%%%%%%%%%%%%%%%%%%%%%%%%%%%%%%%%%%
\vskip .4cm\noindent{\bf electron--$G^\circ_\mu$ vertex}:
\be\label{e-g0} ie\tan{\delta_{13}}\exp{(ip\times k)}\gamma^\mu,\ee

%%%%%%%%%%%%%%%%%%%%%%%%%%%%%%%%%%%%%%%%%%%%%%%%%%%%%%%%%%%%%%%%%%%%%%%%%%%%%%%%%%%%%%%%%%%%%%%%%%%%%%%%%%%%%%
 \vskip .4cm\noindent{\bf down
quark--$G^\circ_\mu$ vertex}: \be \label{d-g0}-\frac 1 3
ie\cot\delta_{13}\exp{(ip\times k)}\gamma^\mu,\ee

%%%%%%%%%%%%%%%%%%%%%%%%%%%%%%%%%%%%%%%%%%%%%%%%%%%%%%%%%%%%%%%%%%%%%%%%%%%%%%%%%%%%%%%%%%%%%%%%%%%%%%%%%%%%%%%%%%%%
 \vskip .4cm\noindent{\bf up
quark--$G^\circ_\mu$ vertex}: \be
\label{u-g0}-ie\{(\tan\delta_{13}-\frac 1 3
\cot\delta_{13})\exp{(ik\times
p)}-\frac{2i}{3}\cot\delta_{13}\sin(k\times p)\}\gamma^\mu,\ee

%%%%%%%%%%%%%%%%%%%%%%%%%%%%%%%%%%%%%%%%%%%%%%%%%%%%%%%%%%%%%%%%%%%%%%%%%%%%%%%%%%%%%%%%%%%%%%%%%%%%%%%%%%%%%%%%%%%%%%%%%%%%
 \vskip .4cm\noindent{\bf neutrino--photon vertex}:
 \be \label{n-p}2e\sin{(ip\times k)}\gamma^\mu,\ee

 %%%%%%%%%%%%%%%%%%%%%%%%%%%%%%%%%%%%%%%%%%%%%%%%%%%%%%%%%%%%%%%%%%%%%%%%%%%%%%%%%%%%%%%%%%%%%%%%%%%%%%%%%%%%%%%%%%%%%%%%%%%%%%%%%%%%%555
  \vskip .4cm\noindent{\bf neutrino--$G^\circ_\mu$ vertex}: \be\label{n-g0} -2e\tan\delta_{13}\sin{(ip\times k)}\gamma^\mu,\ee

%%%%%%%%%%%%%%%%%%%%%%%%%%%%%%%%%%%%%%%%%%%%%%%%%%%%%%%%%%%%%%%%%%%%%%%%%%%%%%%%%%%%%%%%%%%%%%%%%%%%%%%%%%%%%%%%%%%%%%%%%%%%%%%%%%%%%%%%%%
 \vskip .4cm\noindent{\bf up and down
quark--gluon vertex}: \be \label{u&d-g}-\frac 1 2 ig_3\exp{(ip\times
k)}\gamma^\mu T^a,\ee
%%%%%%%%%%%%%%%%%%%%%%%%%%%%%%%%%%%%%%%%%%%%%%%%%%%%%%%%%%%%%%%%%%%%%%%%%%%%%%%%%%%%%%%%%%%%%%
where $p$ and $k$ are the matter field and gauge field,
respectively. Also we have defined $p\times k=\frac 1 2
\th^{\mu\nu}p_\mu k_\nu$.

 \vskip
2cm \noindent{\bf gluon propagator}. \be -\frac {i}{p^2}
\delta_{ab} g_{\mu\nu}\ee \vskip .2cm \noindent{\bf photon and
$G^{\circ}_{\mu}$ propagator}. \be -\frac {i}{p^2} g_{\mu\nu}\ee
\vskip .2cm \noindent{\bf ghost(photon and $G^{\circ}_{\mu}$)
propagator}. \be -\frac {i}{p^2} g_{\mu\nu}\ee
 \vskip .2cm\noindent{\bf ghost(gluon) propagator}.
\be -\frac {i}{p^2} \delta_{ab} g_{\mu\nu}\ee

\vskip .4cm \noindent{\bf 3--gluon vertex}.
%%%%%%%%%%%%%%%%%%%%%%%%%%%%%%%%%%%%%%
\begin{figure}
\centerline{\epsfysize=2in\epsfxsize=3in\epsffile{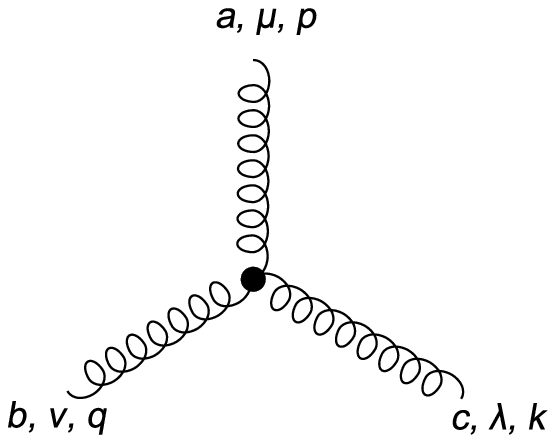}}\caption{3--gluon
 }
\label{fig5}
\end{figure}
%%%%%%%%%%%%%%%%%%%%%%%%%%%%%%%%%%%%%%%%%%%%%%%%%%%%%%

\be\label{3g}-g_3{\sin\delta_{13}}(f_{abc}\,\cos (p\times q) +
d_{abc}\,\sin (p\times q))I \ee

\vskip .4cm \noindent{\bf 3--photon vertex}.
%%%%%%%%%%%%%%%%%%%%%%%%%%%%%%%%%%%%%%
\begin{figure}[b]
\centerline{\epsfysize=2in\epsfxsize=3in\epsffile{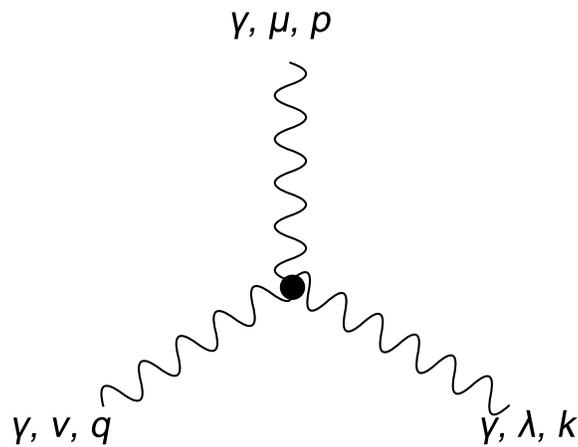}}\caption{3--photon
 }
\label{fig6}
\end{figure}
%%%%%%%%%%%%%%%%%%%%%%%%%%%%%%%%%%%%%%%%%%%%%%%%%%%%%%

\be \label{3p}-\frac{2}{3}e(1+ 2\cos^{2}\delta_{13})\sin (p\times
q)I \ee

\vskip .4cm \noindent{\bf 3--$G^{\circ}$ vertex}.
%%%%%%%%%%%%%%%%%%%%%%%%%%%%%%%%%%%%%%
\begin{figure}
\centerline{\epsfysize=2in\epsfxsize=3in\epsffile{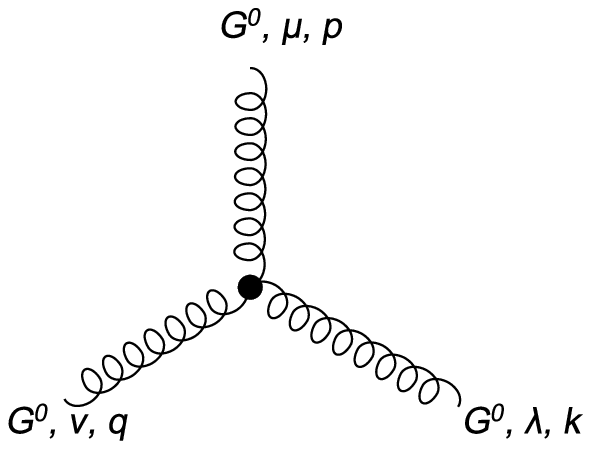}}\caption{3--$G^{\circ}$
 }
\label{fig7}
\end{figure}
%%%%%%%%%%%%%%%%%%%%%%%%%%%%%%%%%%%%%%%%%%%%%%%%%%%%%%

 \be \label{3g0}e(2\tan\delta_{13}\sin^{2}\delta_{13} -
\frac{2}{3}\cot\delta_{13}\cos^{2}\delta_{13})\sin (p\times q)I \ee

\vskip .4cm\noindent{\bf photon--two gluons vertex}.
%%%%%%%%%%%%%%%%%%%%%%%%%%%%%%%%%%%%%%
\begin{figure}[b]
\centerline{\epsfysize=2in\epsfxsize=3in\epsffile{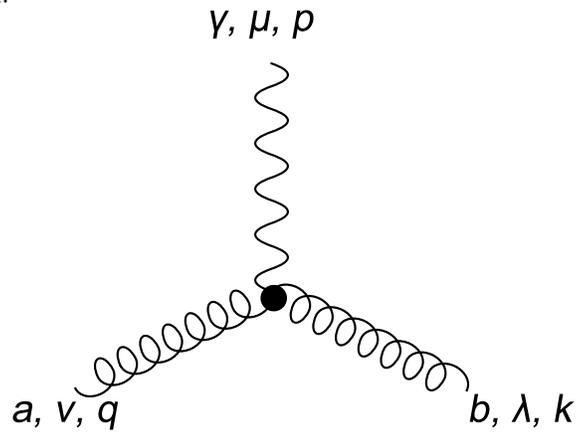}}\caption{photon
- gluon
 }
\label{fig8}
\end{figure}
%%%%%%%%%%%%%%%%%%%%%%%%%%%%%%%%%%%%%%%%%%%%%%%%%%%%%%

 \be\label{p-2g}-e\frac{2}{3}\delta_{ab}\sin (p\times q)I \ee

\vskip .4cm \noindent{\bf $G^{\circ}$--two gluons vertex}.
%%%%%%%%%%%%%%%%%%%%%%%%%%%%%%%%%%%%%%
\begin{figure}
\centerline{\epsfysize=2in\epsfxsize=3in\epsffile{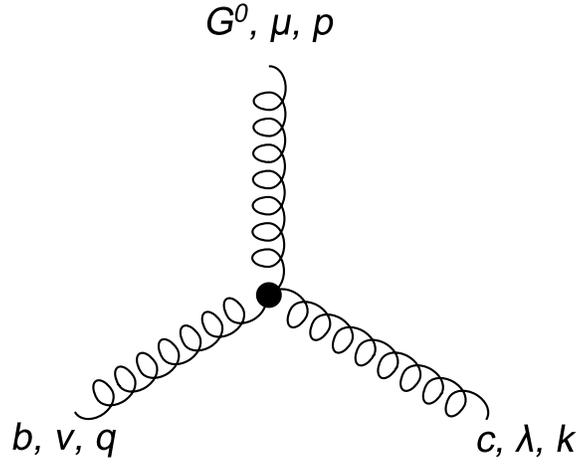}}\caption{$G^{\circ}$
- gluon
 }
\label{fig9}
\end{figure}
%%%%%%%%%%%%%%%%%%%%%%%%%%%%%%%%%%%%%%%%%%%%%%%%%%%%%%

\be \label{g0-2g}-e\frac{2}{3}\delta_{ab}\cot\delta_{13}\sin
(p\times q)I\ee

\vskip .4cm \noindent{\bf $G^{\circ}$--two photons vertex}.
%%%%%%%%%%%%%%%%%%%%%%%%%%%%%%%%%%%%%%
\begin{figure}[b]
\centerline{\epsfysize=2in\epsfxsize=3in\epsffile{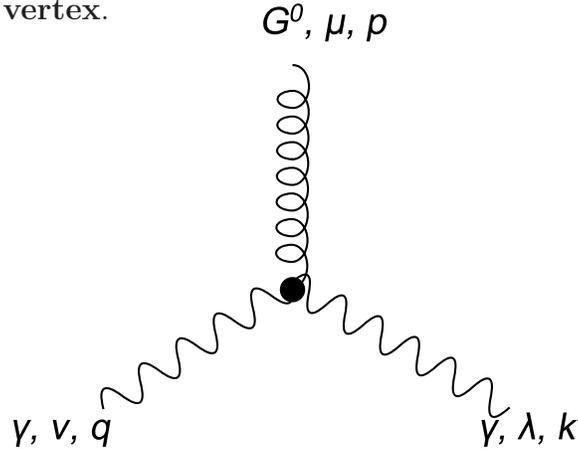}}\caption{$G^{\circ}$
- two photons
 }
\label{fig10}
\end{figure}
%%%%%%%%%%%%%%%%%%%%%%%%%%%%%%%%%%%%%%%%%%%%%%%%%%%%%%
 \be\label{g0-2p}-\frac{2}{3}e\sin2\delta_{13}\sin (p\times q)I\ee

\vskip .4cm \noindent{\bf photon--two $G^{\circ}$  vertex}.

%%%%%%%%%%%%%%%%%%%%%%%%%%%%%%%%%%%%%%
\begin{figure}
\centerline{\epsfysize=2in\epsfxsize=3in\epsffile{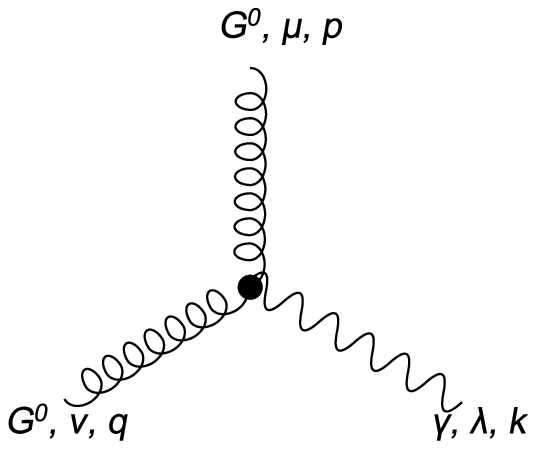}}\caption{two
$G^{\circ}$ - photon
 }
\label{fig11}
\end{figure}
%%%%%%%%%%%%%%%%%%%%%%%%%%%%%%%%%%%%%%%%%%%%%%%%%%%%%%
\be\label{p-2g0}-\frac{2}{3}e(1+ 2\sin^{2}\delta_{13})\sin (p\times
q)I \ee
 Where
$I$ is : \be I =(g_{\mu\nu}\,(p-q)_\lambda +
g_{\nu\lambda}\,(q-k)_\mu+\,g_{\lambda\mu}(k-p))\ee

\vskip .4cm \noindent{\bf photon--three gluons vertex}.
%%%%%%%%%%%%%%%%%%%%%%%%%%%%%%%%%%%%%%
\begin{figure}
\centerline{\epsfysize=2in\epsfxsize=3in\epsffile{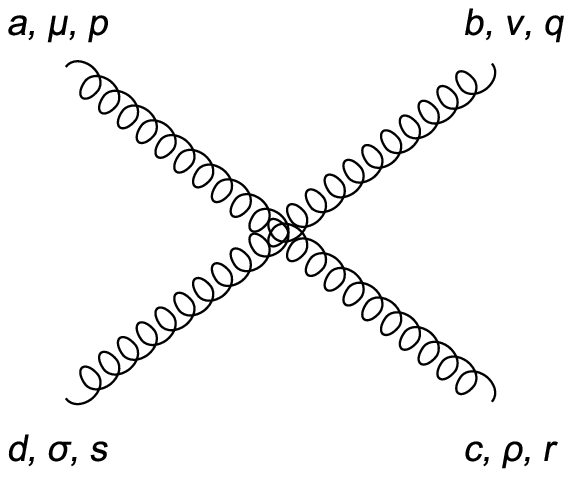}}\caption{$G^{\circ}$
- gluon } \label{fig12}
\end{figure}
%%%%%%%%%%%%%%%%%%%%%%%%%%%%%%%%%%%%%%%%%%%%%%%%%%%%%%
 \be\label{p-3g}-\frac{8ie^{2}}{3\sin\delta_{13}}I_1 \ee

\vskip .4cm \noindent{\bf $G^{\circ}$--three gluons vertex}.
%%%%%%%%%%%%%%%%%%%%%%%%%%%%%%%%%%%%%%
\begin{figure}
\centerline{\epsfysize=2in\epsfxsize=3in\epsffile{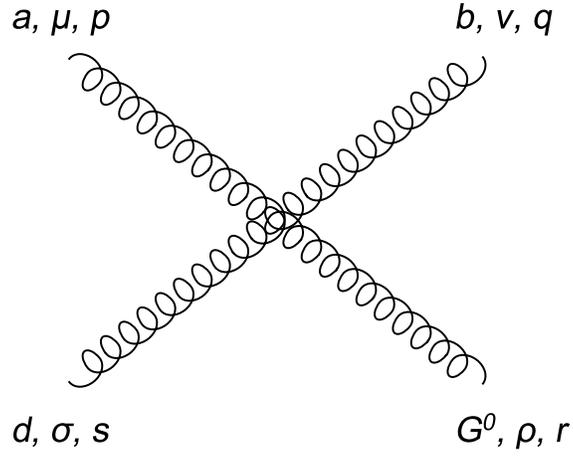}}\caption{$G^{\circ}$
- three gluons (photon - three gluons)
 }
\label{fig13}
\end{figure}
%%%%%%%%%%%%%%%%%%%%%%%%%%%%%%%%%%%%%%%%%%%%%%%%%%%%%%
\be\label{g0-3g}-\frac{8ie^{2}\cot\delta_{13}}{3\sin\delta_{13}}I_1
\ee

\vskip .4cm \noindent{\bf two $G^{\circ}$--two gluons vertex}.
%%%%%%%%%%%%%%%%%%%%%%%%%%%%%%%%%%%%%%
\begin{figure}[b]
\centerline{\epsfysize=2in\epsfxsize=3in\epsffile{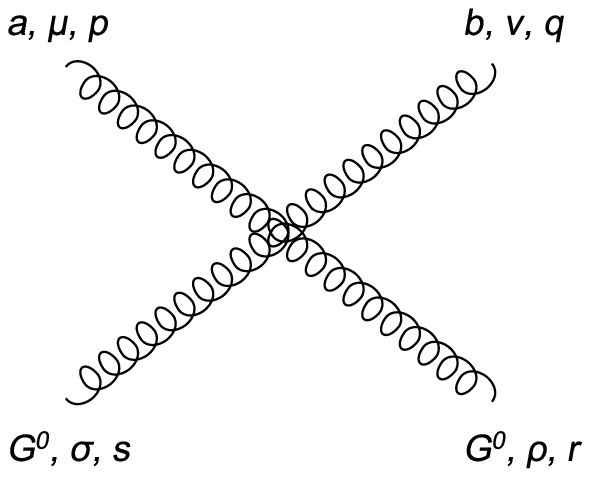}}\caption{two
$G^{\circ}$ - two gluon
 }
\label{fig14}
\end{figure}
%%%%%%%%%%%%%%%%%%%%%%%%%%%%%%%%%%%%%%%%%%%%%%%%%%%%%%
\be\label{2g0-2g}-2i(\frac{2e}{3})^{2}\delta_{ab}\cot^{2}\delta_{13}I_2
\ee
 \vskip .4cm \noindent{\bf two photon--two gluons vertex}.
%%%%%%%%%%%%%%%%%%%%%%%%%%%%%%%%%%%%%%
\begin{figure}
\centerline{\epsfysize=2in\epsfxsize=3in\epsffile{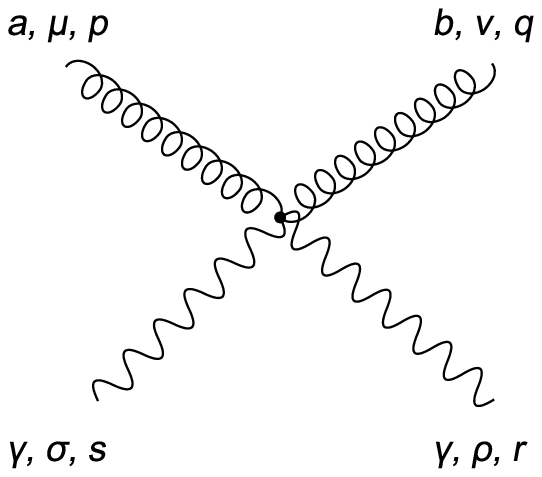}}\caption{two
photon - two gluon
 }
\label{fig15}
\end{figure}
%%%%%%%%%%%%%%%%%%%%%%%%%%%%%%%%%%%%%%%%%%%%%%%%%%%%%%
\be\label{2p-2g}-2i(\frac{2e}{3})^{2}\delta_{ab}I_2 \ee

\vskip .4cm \noindent{\bf 4-- photon vertex}.
%%%%%%%%%%%%%%%%%%%%%%%%%%%%%%%%%%%%%%%%%
\begin{figure}
\centerline{\epsfysize=2in\epsfxsize=3in\epsffile{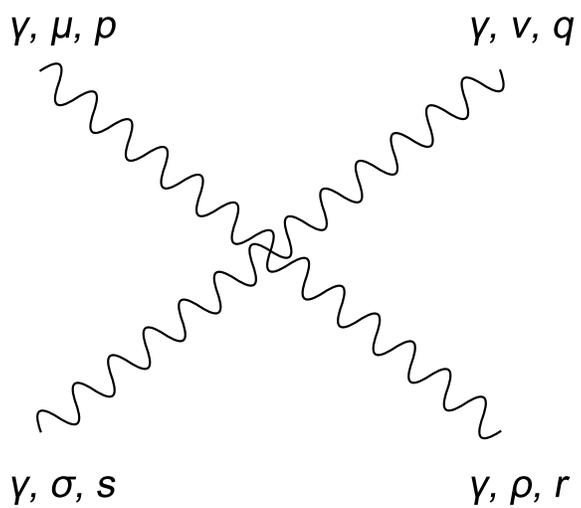}}\caption{4--
photon } \label{fig16}
\end{figure}
%%%%%%%%%%%%%%%%%%%%%%%%%%%%%%%%%%%%%%%%%%%%%%%
\be \label{4p}-i(\frac{2e}{3})^{2}(1 + 8\cos^{2}\delta_{13})I_2 \ee

\vskip .4cm \noindent{\bf 4-- $G^{\circ}$ vertex}.
%%%%%%%%%%%%%%%%%%%%%%%%%%%%%%%%%%%%%%%%%
\begin{figure}
\centerline{\epsfysize=2in\epsfxsize=3in\epsffile{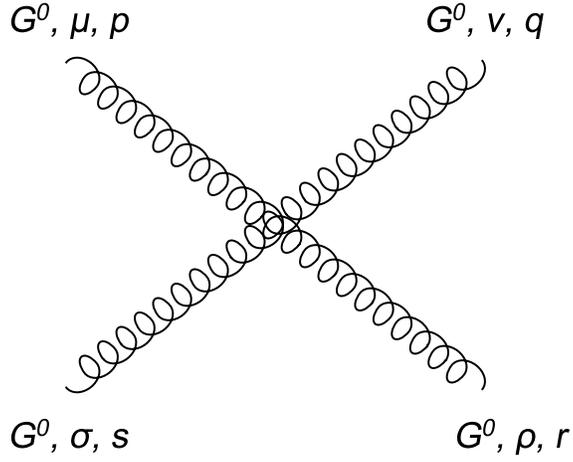}}\caption{4--
$G^{\circ}$ } \label{fig17}
\end{figure}
%%%%%%%%%%%%%%%%%%%%%%%%%%%%%%%%%%%%%%%%%%%%%%%
\be\label{4g0}-4ie^{2}(1/9\cot^{2}\delta_{13}\cos^{2}\delta_{13} +
\tan^{2}\delta_{13}\sin^{2}\delta_{13})I_2 \ee \vskip .4cm
\noindent{\bf two photons--two $G^{\circ}$ vertex}.
%%%%%%%%%%%%%%%%%%%%%%%%%%%%%%%%%%%%%%%%%
\begin{figure}[b]
\centerline{\epsfysize=2in\epsfxsize=2.7in\epsffile{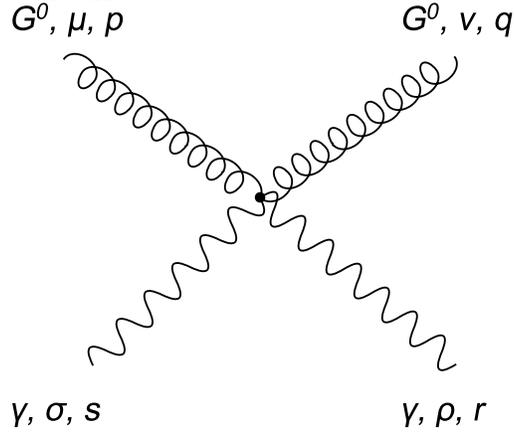}}\caption{two
$G^{\circ}$ - two photon } \label{fig18}
\end{figure}
%%%%%%%%%%%%%%%%%%%%%%%%%%%%%%%%%%%%%%%%%%%%%%%
\be\label{2p-2g0}-i(\frac{2e}{3})^{2}(1 + 8\sin^{2}\delta_{13})I_2
\ee \vskip .1cm \noindent{\bf three photons-- $G^{\circ}$ vertex}.
%%%%%%%%%%%%%%%%%%%%%%%%%%%%%%%%%%%%%%%%%
\begin{figure}
\centerline{\epsfysize=2in\epsfxsize=3in\epsffile{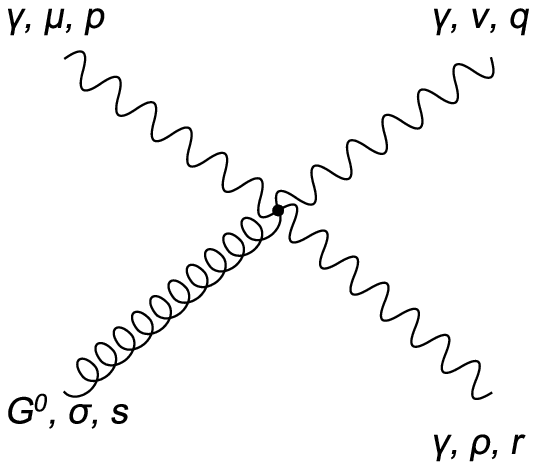}}\caption{
$G^{\circ}$ - three photon } \label{fig19}
\end{figure}
%%%%%%%%%%%%%%%%%%%%%%%%%%%%%%%%%%%%%%%%%%%%%%%
\be\label{3p-g0}i(\frac{4e}{3})^{2}\sin2\delta_{13}I_2 \ee \vskip
.1cm \noindent{\bf photon-- $G^{\circ}$--two gluons vertex}.
%%%%%%%%%%%%%%%%%%%%%%%%%%%%%%%%%%%%%%%%%
\begin{figure}
\centerline{\epsfysize=2in\epsfxsize=3in\epsffile{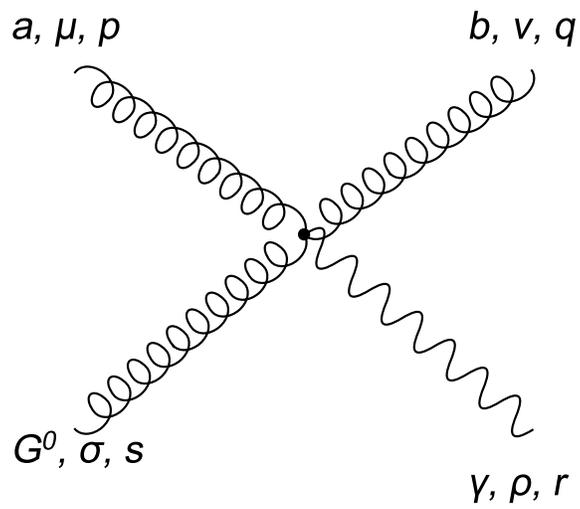}}\caption{
$G^{\circ}$ - photon - two gluon } \label{fig20}
\end{figure}
%%%%%%%%%%%%%%%%%%%%%%%%%%%%%%%%%%%%%%%%%%%%%%%
\be\label{p-g0-2g}-i\frac 8 9 e^{2}\cot\delta_{13}\delta_{ab}I_2 \ee

\vskip .4cm \noindent{\bf photon--three $G^{\circ}$ vertex}.
%%%%%%%%%%%%%%%%%%%%%%%%%%%%%%%%%%%%%%%%%
\begin{figure}
\centerline{\epsfysize=2in\epsfxsize=3in\epsffile{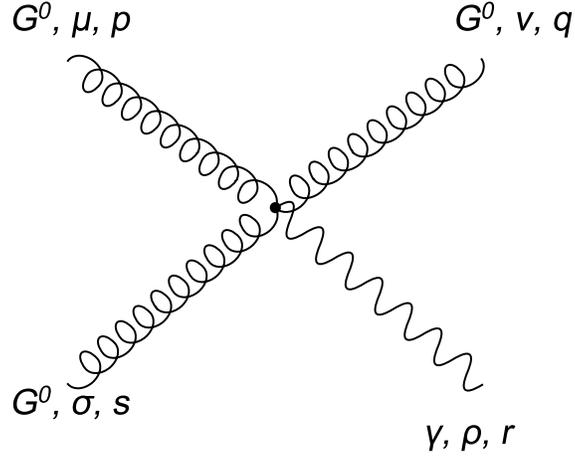}}\caption{three
$G^{\circ}$ - photon } \label{fig21}
\end{figure}
%%%%%%%%%%%%%%%%%%%%%%%%%%%%%%%%%%%%%%%%%%%%%%%

\be\label{p-3g0}-4ie^{2}(\cot\delta_{13}\cos^{2}\delta_{13} -
\tan\delta_{13}\sin^{2}\delta_{13})I_2 \ee

Where $I_1 , I_2$ are

\bea I_1 = \sqrt{\frac 2 3}[&&\hspace{-10mm}(f_{abc}\,\cos (p\times
q) +
d_{abc}\,\sin (p\times q))\no\\
&&\cdot\,\sin (r\times s)
(g_{\mu\rho}\,g_{\nu\sigma}- g_{\mu\sigma}\,g_{\nu\rho})\no\\
\hspace{-3mm}&+&\hspace{-3mm}\left(f_{acb}\,\cos (s\times q) + d_{acb}\,\sin (s\times q)\right)\no\\
&&\cdot\,\sin (p\times r)
(g_{\mu\sigma}\,g_{\nu\rho}- g_{\mu\nu}\,g_{\rho\sigma})\no\\
\hspace{-3mm}&+&\hspace{-3mm}\left(f_{acb}\,\cos (p\times s) + d_{acb}\,\sin (p\times s)\right)\no\\
&&\cdot \,\sin (q\times r) (g_{\mu\nu}\,g_{\rho\sigma}-
g_{\mu\rho}\,g_{\nu\sigma})]\eea

%\bea I^\prime_1 = -ig^2[&+&(f_{abd}\,\cos (p\times q) +
%d_{abd}\,\sin (p\times q))\no\\
%&&\quad\quad\cdot\,\sin (r\times s)
%(g_{\mu\rho}\,g_{\nu\sigma}- g_{\mu\sigma}\,g_{\nu\rho})\no\\
%&+&\left(f_{abd}\,\cos (p\times r) + d_{abd}\,\sin (p\times r)\right)\no\\
%&&\quad\quad\cdot\,\sin (s\times q)
%(g_{\mu\sigma}\,g_{\nu\rho}- g_{\mu\nu}\,g_{\rho\sigma})\no\\
%&+&\left(f_{abd}\,\cos (p\times s) + d_{abd}\,\sin (p\times s)\right)\no\\
%&&\quad\quad\cdot \,\sin (q\times r) (g_{\mu\nu}\,g_{\rho\sigma}-
%g_{\mu\rho}\,g_{\nu\sigma})]\eea

\bea I_2 = [&&\hspace{-10mm} \,\sin (p\times q)\sin (r\times s)
(g_{\mu\rho}\,g_{\nu\sigma}- g_{\mu\sigma}\,g_{\nu\rho})\no\\
\hspace{-3mm}&+&\hspace{-3mm}\,\sin (p\times r)\sin (s\times q)
(g_{\mu\sigma}\,g_{\nu\rho}- g_{\mu\nu}\,g_{\rho\sigma})\no\\
\hspace{-3mm}&+&\hspace{-3mm}\,\sin (p\times s)\sin (q\times
r)(g_{\mu\nu}\,g_{\rho\sigma}- g_{\mu\rho}\,g_{\nu\sigma})] \eea

\vskip .4cm \noindent{\bf photon--ghosts(photon) vertex}
%%%%%%%%%%%%%%%%%%%%%%%%%%%%%%%%%%%%%%%%%
\begin{figure}
\centerline{\epsfysize=2in\epsfxsize=3in\epsffile{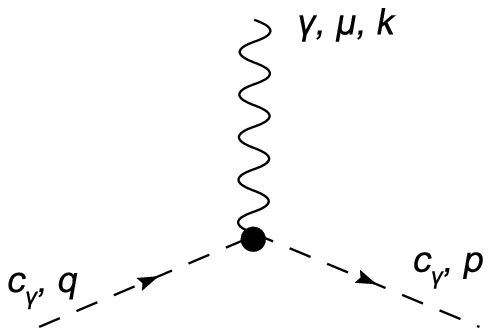}}\caption{photon--ghosts(photon)
} \label{fig22}
\end{figure}
%%%%%%%%%%%%%%%%%%%%%%%%%%%%%%%%%%%%%%%%%%%%%%%

\be \label{p-gh(p)}\frac{2}{3}e(1 + 2\cos^{2}\delta_{13})\,p_\mu
\sin (p\times q) \ee

\vskip .4cm \noindent{\bf photon--ghosts($G^{\circ}$) vertex}
%%%%%%%%%%%%%%%%%%%%%%%%%%%%%%%%%%%%%%%%%
\begin{figure}
\centerline{\epsfysize=2in\epsfxsize=3in\epsffile{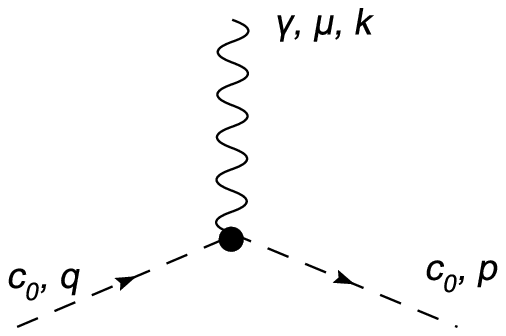}}\caption{photon--ghosts($G^{\circ}$)
} \label{fig23}
\end{figure}
%%%%%%%%%%%%%%%%%%%%%%%%%%%%%%%%%%%%%%%%%%%%%%%
\be \label{p-gh(g0)}\frac{2}{3}e(1 + 2\sin^{2}\delta_{13})\,p_\mu
\sin (p\times q) \ee

\vskip .4cm \noindent{\bf $G^{\circ}$--ghosts($G^{\circ}$) vertex}
%%%%%%%%%%%%%%%%%%%%%%%%%%%%%%%%%%%%%%%%%
\begin{figure}
\centerline{\epsfysize=2in\epsfxsize=3in\epsffile{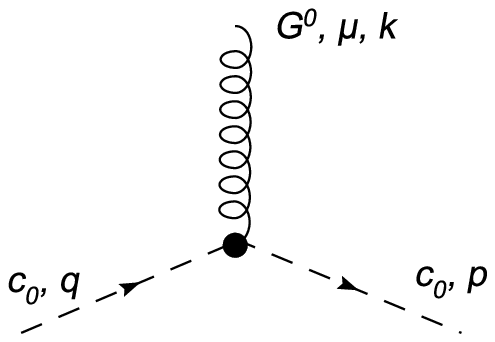}}\caption{$G^{\circ}$--ghosts($G^{\circ}$)
} \label{fig24}
\end{figure}
%%%%%%%%%%%%%%%%%%%%%%%%%%%%%%%%%%%%%%%%%%%%%%%
\be\label{g0-gh(g0)}
-e(\frac{1}{3}\cot\delta_{13}\cos^{2}\delta_{13} -
\tan\delta_{13}\sin^{2}\delta_{13})\,p_\mu \sin (p\times q) \ee

\vskip .4cm \noindent{\bf $G^{\circ}$--ghosts(photon) vertex.}
%%%%%%%%%%%%%%%%%%%%%%%%%%%%%%%%%%%%%%%%%
\begin{figure}[b]
\centerline{\epsfysize=2in\epsfxsize=3in\epsffile{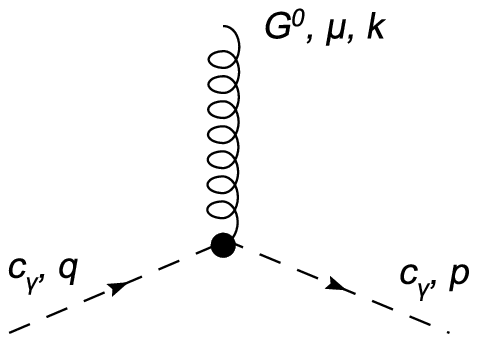}}\caption{$G^{\circ}$--ghosts(photon)
} \label{fig25}
\end{figure}
%%%%%%%%%%%%%%%%%%%%%%%%%%%%%%%%%%%%%%%%%%%%%%%
\be \label{g0-gh(p)}-\frac{2}{3}e\sin2\delta_{13}\,p_\mu \sin
(p\times q) \ee

\vskip .4cm \noindent{\bf photon--ghosts(photon--$G^{\circ}$)
vertex}
%%%%%%%%%%%%%%%%%%%%%%%%%%%%%%%%%%%%%%%%%
\begin{figure}
\centerline{\epsfysize=2in\epsfxsize=3in\epsffile{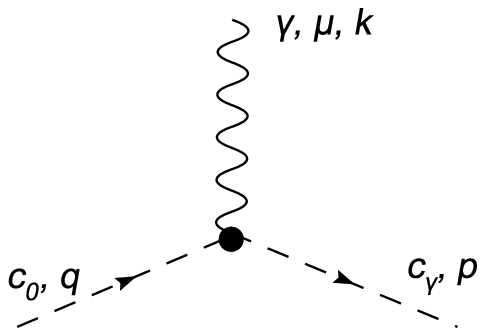}}\caption{photon--ghosts(photon
- $G^{\circ}$) } \label{fig26}
\end{figure}
%%%%%%%%%%%%%%%%%%%%%%%%%%%%%%%%%%%%%%%%%%%%%%%
\be \label{p-gh(p-g0)}-\frac{2}{3}e\sin2\delta_{13}\,p_\mu \sin
(p\times q) \ee

\vskip .4cm \noindent{\bf $G^{\circ}$--ghosts(photon--$G^{\circ}$)
vertex}
%%%%%%%%%%%%%%%%%%%%%%%%%%%%%%%%%%%%%%%%%
\begin{figure}[b]
\centerline{\epsfysize=2in\epsfxsize=3in\epsffile{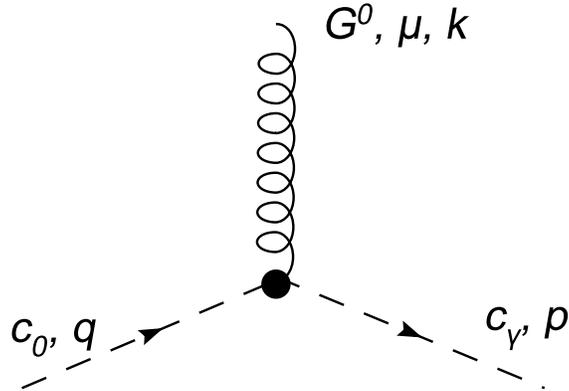}}\caption{$G^{\circ}$--ghosts(photon
- $G^{\circ}$) } \label{fig27}
\end{figure}
%%%%%%%%%%%%%%%%%%%%%%%%%%%%%%%%%%%%%%%%%%%%%%%
\be \label{g0-gh(p-g0)}\frac{2}{3}e(1 + 2\sin^{2}\delta_{13})\,p_\mu
\sin (p\times q) \ee

\vskip .5cm \noindent{\bf photon--ghosts(gluon) vertex}
%%%%%%%%%%%%%%%%%%%%%%%%%%%%%%%%%%%%%%%%%
\begin{figure}
\centerline{\epsfysize=2in\epsfxsize=3in\epsffile{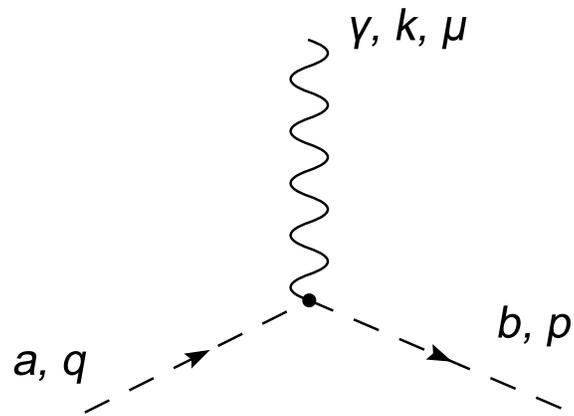}}\caption{photon--ghosts(gluon)
} \label{fig28}
\end{figure}
%%%%%%%%%%%%%%%%%%%%%%%%%%%%%%%%%%%%%%%%%%%%%%%
\be \label{p-gh(g)}\frac{2}{3}e\delta_{ab}\,p_\mu \sin (p\times q)
\ee

\vskip .3cm \noindent{\bf $G^{\circ}$--ghosts(gluon) vertex}
%%%%%%%%%%%%%%%%%%%%%%%%%%%%%%%%%%%%%%%%%
\begin{figure}
\centerline{\epsfysize=2in\epsfxsize=3in\epsffile{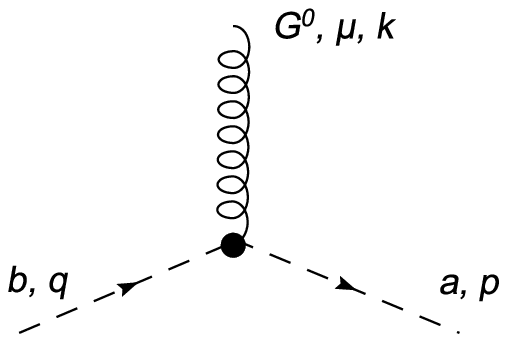}}\caption{$G^{\circ}$--ghosts(gluon)
} \label{fig29}
\end{figure}
%%%%%%%%%%%%%%%%%%%%%%%%%%%%%%%%%%%%%%%%%%%%%%%
\be \label{g0-gh(g)}-\frac{2}{3}e\delta_{ab}\cot\delta_{13}\,p_\mu
\sin (p\times q) \ee

\vskip .3cm \noindent{\bf gluon--ghosts(gluon)
vertex}\be\label{g-gh(g)} -g_3\,p_\mu (f_{abc}\,\cos (p\times q) -
d_{abc}\,\sin (p\times q)) \ee

%%%%%%%%%%%%%%%%%%%
\begin{figure}
\centerline{\epsfysize=2in\epsfxsize=3in\epsffile{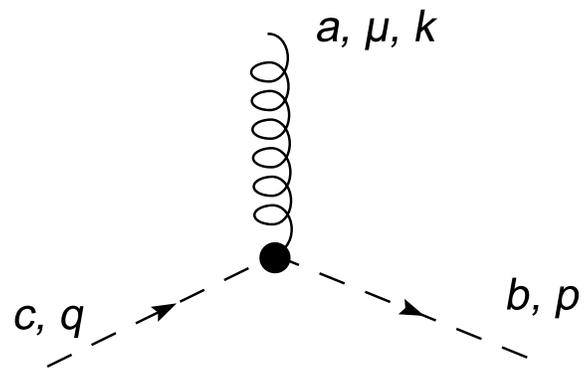}}\caption{gluon
- ghost(gluon)
 }
\label{fig30}
\end{figure}
\newpage

%=========================================================================

\end{document}